# A Century of Evolution in the Complexity of the United States Legal Code

Dawoon Jeong[1,2*], James Holehouse[3,*†], Jisung Yoon[4], Christopher P. Kempes[3], Geoffrey B. West[3], and Hyejin Youn[2,3,5,‡]

[1]Kellogg School of Management, Northwestern University, Evanston, IL, USA

[2]Northwestern Institute on Complex Systems, Evanston, IL, USA

[3]Santa Fe Institute, Santa Fe, NM, USA.

[4]KDI School of Public Policy and Management, South Korea

[5]Graduate School of Business, Seoul National University, Seoul, South Korea

[*]These authors contributed equally to this work.

[†]jamesholehouse1@gmail.com

[‡]h.youn@snu.ac.kr

July 18, 2025

## Abstract

As societies confront increasingly complex regulatory demands in domains such as digital governance, climate policy, and public health, there is a pressing need to understand how legal systems evolve, where they concentrate regulatory attention, and how their institutional architectures shape capacity for adaptation. Yet, the long-term structural dynamics of law remain empirically underexplored. Here, we provide a versioned, machine-readable record of the United States Code (U.S. Code), the primary compilation of federal statutory law in the United States, covering the entire history of the Code from 1926 to 2023. We include not only the curated text in Code but also its structural and linguistic complexity: word counts, vocabulary statistics, hierarchical organization (titles, chapters, sections, subsections), and cross-references among titles. In this way, the dataset offers an empirical foundation for large-scale and long-term interdisciplinary analysis of the growth, reorganization, and internal logic of statutory systems. The dataset is released on GitHub with comprehensive documentation to support reuse across legal studies, data science, complexity research, and institutional analysis.

# Background & Summary

Understanding how legal frameworks respond to or shape societal change poses both theoretical and empirical challenges that cut across disciplines, legal scholars, historians, social scientists, economists, and policymakers [1, 2, 3, 4, 5, 6, 7, 8, 9]. Within this broader question, the *U.S. Code*—the official compilation of federal statutory law—serves as a valuable empirical record: it not only codifies congressional governance but also captures the imprint of long-term shifts in American society, the economy, and technology. Building on this foundation, constructing a comprehensive dataset of the Code opens the door to systematic analysis of the structure, complexity, and temporal dynamics of law and governance.

First codified in 1926, the U.S. Code offers a structured record of how federal law adapts to shifting social values, economic priorities, and technological change [5]. This historical evolution provides an empirical foundation for examining how legal statutes respond to major societal changes; how past laws shape future legislation; and how cross-referencing between internal aspects of the Code increases legal complexity [10, 5]. The Code expands and is periodically reorganized in response to institutional change and broader political and normative shifts, leaving enduring imprints major policy domains, and corresponding legal domains such as civil rights, environmental regulation, consumer protection, and digital privacy. For example, following the 9/11 attacks, Title 6, originally titled "Surety Bonds", was recodified as "Domestic Security", and the original content was moved to a chapter of Title 31, which now covers "Money and Finance."Such shifts show how written rules and laws serve not only as a tool of governance but also as a living archive of societal priorities and institutional transformation [6].

Despite its central role in shaping and reflecting society, long-term historical analysis of the U.S. Code has remained limited, largely due to the limited access to the old texts. Here, we contribute to the field by availing a comprehensive data set that spans nearly the entire history of the U.S. Code (first conceived in 1926). The challenge is to clean and reconstruct early editions that exist only in printed form for systematic, computational analysis of the Code's evolution in both content and structure. We first employ Generative AI-based text pre-processing to accurately convert Optical Character Recognition (OCR) text into structured digital formats, bridging the analog with current digital datasets (see Methods). We provide statistical tests for methodological validations to ensure the recovered legislative records enable unified, computational analysis of the Code's evolution in both content and structure across nearly a century. In addition to the full text, we provide a structured dataset such that researchers analyze fine-grained changes in legal architecture for future empirical research toward regulatory evolution, legal codification, and the dynamics of institutional knowledge.

Figure 1 illustrates our dataset, organized into three levels of legal structure: (i) the textual content of each title, (ii) the internal structural hierarchy within titles, and (iii) the cross-reference network linking different titles. To streamline, we refer to these as Dataset 1, Dataset 2, and Dataset 3, respectively.

Dataset 1 provides the data at the most granular level: the textual content, vocabulary, and linguistic composition of legal provisions. The structural statistics include word counts, vocabulary size (i.e., the number of unique words), frequency distributions, and concentration metrics such as Shannon entropy [2]. For example, the emergence of terms like *cybersecurity*, *data breach*, and *genetic information* in the 2000s marks the incorporation of new technological and societal concerns into statutory language [11, 5]. Therefore, these statistics provide overall structural comparisons across legal domains by quantifying the rate at which new vocabulary emerges relative to total word count—such as Heaps' law of a corpus of legal text. For instance, Title 3 (The President) exhibits faster vocabulary growth than Title 19 (Customs Duties), as captured by their respective Heaps' law exponents. We further provide word frequency distributions $f_{i,k,t}$—the frequency of word $i$ in title $k$ at time $t$—to assess lexical richness and semantic drift over time [12].

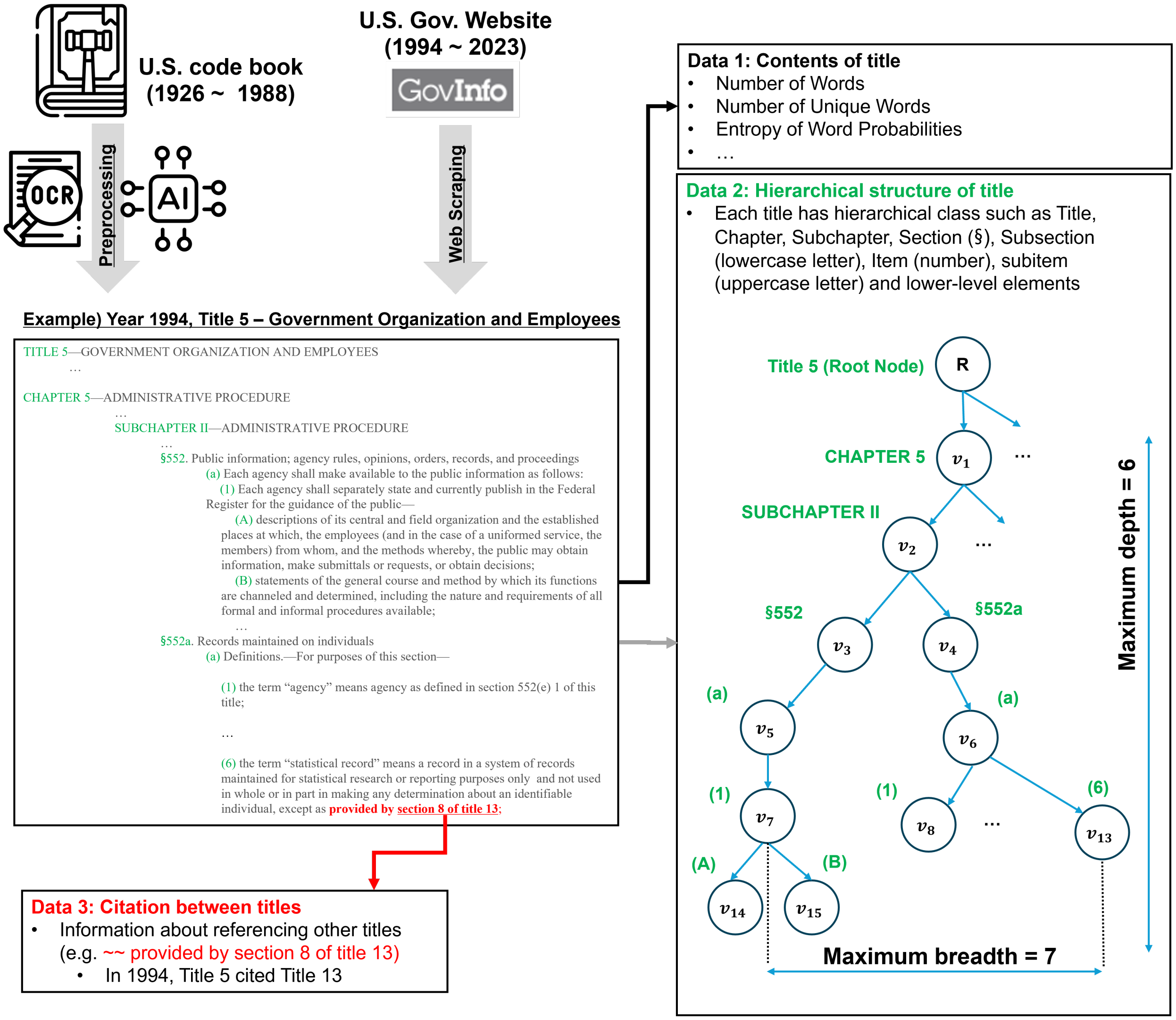

**Figure 1: Overview of U.S. Code data processing and structure.** This figure illustrates the process of extracting and analyzing hierarchical legal text from the U.S. Code. Data sources include scanned volumes from `HeinOnline` (1926–1988) and digital records from the GovInfo website (1994–2023). OCR images from historical volumes are processed into digitized text using a customized generative AI pipeline with manual reviews (see Methods). From the curated text, we create three datasets: (1) textual content of titles, including word counts, vocabulary size, and word entropy (occurrence frequency distribution); (2) hierarchical structure, where legal text is organized into subdivisions such as *titles*, *chapters*, *subchapters*, *sections* (§), *subsections* (lowercase letters), *items* (numbers), and *subitems* (uppercase letters); and (3) a cross-reference network across different titles.

Beyond individual lexical occurrences (Dataset 1), we provide the internal hierarchical struc-ture of the Code. As with many formal documents, the U.S. Code is organized as a nested system: titles contain chapters, which contain subchapters, sections, and items. Dataset 2 provides this second-order structure that quantifies how legal content is organized across its scope. Some titles are structurally shallow but broad, composed of many top-level chapters with few sections or items; others are narrow but deep, containing numerous sections and multiple layers of nested subdivisions. For example, Title 3 (The President) exhibits deep internal structure with numerous subchapters and items, while Title 36 (Patriotic and National Observances) is relatively flat. Using both natural language processing and customized rule-based parsing, we parse the text to construct these second-order structures for each title, allowing us to measure both the breadth and depth of each title's tree structure. This, in turn, enables comparisons of structural complexity across legal domains and over time.

Finally, Dataset 3 provides the relational level legal architecture, a directed network of cross-references among titles, capturing legal and institutional interdependencies [10, 13, 14]. Titles are not isolated legal domains; many rely on legal concepts codified elsewhere, and their provisions must align with broader statutory logic. As society shifts its priorities, these interdependencies vary across legal domains. Some titles function as hubs—frequently cited by others—while others remain peripheral or are embedded within tightly coupled clusters. For example, Title 26 (Internal Revenue) and Title 7 (Agriculture) are frequently referenced by Title 15 (Commerce and Trade), illustrating the regulatory entwinement of trade, taxation, and agricultural policy. This reference network, therefore, reveals which areas of law are structurally central to the legal system and which operate in relative isolation. We apply network analysis to map these dependencies, identifying central, peripheral, and modular regions within the U.S. Code.

Together, these three structural layers—textual, hierarchical, and relational—provide an integrated perspective on how the U.S. Code has evolved in terms of substance, structure, and interdependence [2, 15, 3, 10, 5, 16, 17]. This dataset builds on two key prior efforts: (1) Li *et al.* (2015) [2], which used unprocessed OCR legal data from `HeinOnline` to explore the evolution of the Code from 1926 to 2010; and (2) Katz *et al.* (2020) [5], which provides 25 years of analysis of the U.S. Code from 1994 to 2018. Building on them, we provide publicly available, fully digitized texts spanning the entire history of the U.S. Code—from 1926 through 2023—while maintaining methodological consistency with the post-1994 digital records. In doing so, our contribution complements existing work and introduces a century-scale resource for systematic, computational analysis of federal statutory law to support robust and scalable research into the dynamics of statutory evolution, legal citation networks, and long-term legislative change.

# Methods

## Data curation for the U.S. Code

The U.S. Code was first published in 1926, followed by a second official edition in 1934. Since then, new editions have been released at six-year intervals. However, the earliest editions were not born digital; their content exists only as scanned images, which are unsuitable for computational analysis. Our first challenge, therefore, is to reconstruct usable and machine-readable text from these scanned sources [18, 19].

We begin with OCR-processed machine-readable text derived from the scanned images of his-torical volumes. As shown in Fig. 2 (a), these OCR outputs, especially for the earlier editions, unfortunately, contain substantial noise—typographical errors, inconsistent spacing, and illegible characters—making it unsuitable for any meaningful analysis without correction. Fig. 2 shows how we clean and reconstruct the text using a generative AI model (Google Gemini 1.5 Flash), with customizing and fine-tuning prompts. This prompt engineering explicitly constrains the model's behavior to recovery tasks only, avoiding any additions or rephrasing beyond the original text.

We validate this curation process in three complementary ways. First, we manually review samples from each title. This review involves iteratively refining the AI prompts to improve output fidelity (see Methods for details of our working prompts). Second, we compare structural and linguistic statistics from our reconstructed texts of the history record (Dataset 1) with those of more recent digitized records (1994 onwards) in the U.S. GovInfo website, confirming consistency across the OCR-derived and digitally born corpora. This method assumes that there is a consistent underlying dynamic in legal texts such that continuations can be found. Third, we benchmark our reconstructed outputs (Datasets 1–3) against published empirical findings from prior studies [2, 5], observing alignment with known patterns (see Methods and SI).

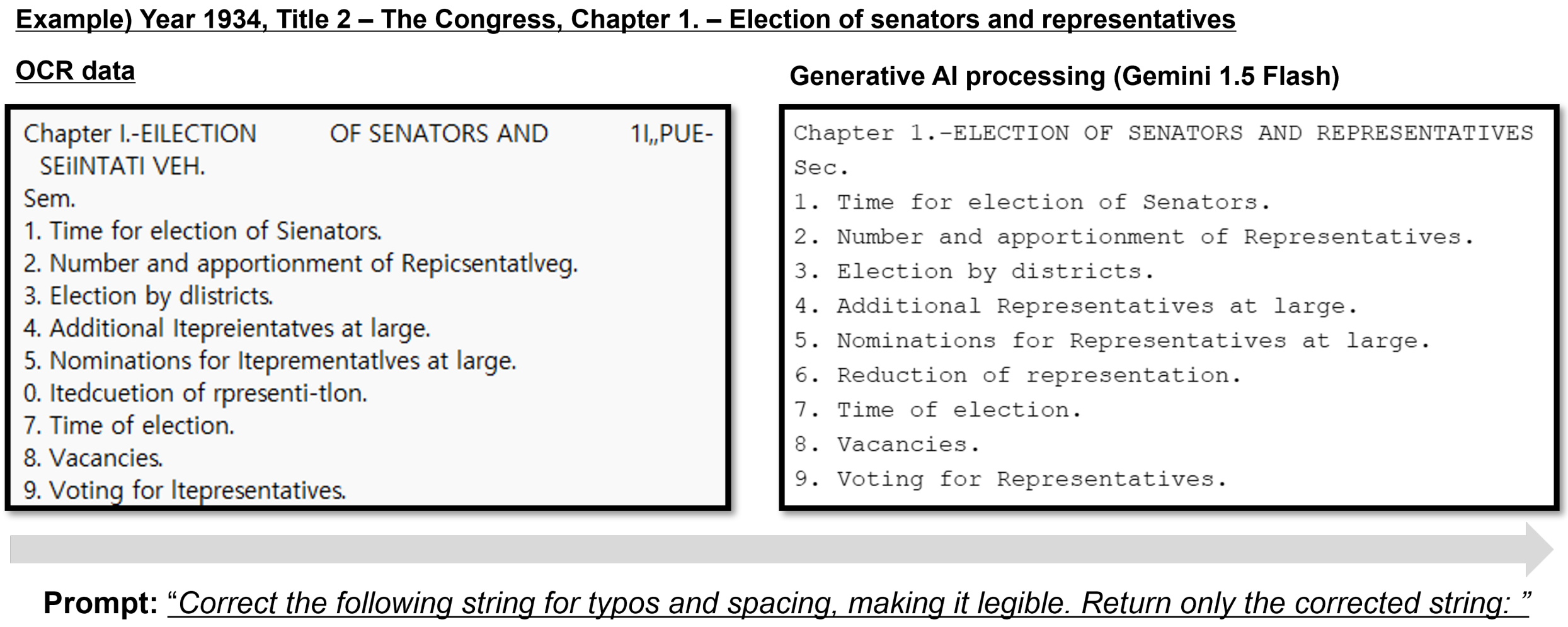

**Figure 2: Generative AI-based preprocessing of OCR text.** The left panel displays raw OCR output from Title 2—The Congress (1934), which includes recognition errors such as misspelled words (e.g., "Sienators" instead of "Senators") and distorted section titles. The right panel presents the corrected version after applying a generative AI model with the prompt: *"Correct the following string for typos and spacing, making it legible. Return only the corrected string."* This AI-based preprocessing step improves the legibility and structural fidelity of OCR-extracted legal text.

Manual review of sample outputs includes comparison between pre- and post-processed texts to ensure accuracy and make necessary revisions as shown in Fig. 2. As a result, we digitize the U.S. Code editions published in 1926, 1934, 1940, 1946, 1952, 1958, 1964, 1970, 1976, 1982, 1988, 1994, and 2000. Note that the 1926 edition contained a high volume of unresolved typographical errors. Therefore, we include the released dataset for the sake of completeness and to support future curation efforts, should improved source materials or technological advances become available. However, we exclude it for the second validation process.

Once the pre-digital dataset is available, we collect the post-digital dataset to be combined. We first code an in-house web crawler to collect the text files from the official *U.S. GovInfo* website for the period from 1994 to 2023 (post-digital). This segment includes both the main editions, published every six years, and the annual supplementary editions, offering a more fine-grained perspective on the evolution of statutory law. Even then, there are a few years of missing downloadable datasets in the *U.S. GovInfo* archive, and thus we interpolated them with the nearest year's dataset for statistical analysis for the second data validity. For example, Titles 7, 28, and 39 in 1994 are unavailable and thus interpolated with their 1995 counterparts. Conversely, Title 18 in 1995 was interpolated with the 1994 version. Additional interpolations were implemented as follows: Title 10 in 2002 used the 2001 edition; Title 46 in both 2004 and 2005 relied on the 2003 version; and Title 50 in 2003 was substituted with the 2002 edition. In 2010, Titles 12, 13, and 14 were supplemented using their respective 2009 versions. Similarly, Title 26 in 2004, Title 5 in 2001, Title 11 in 1998, and Title 10 in 2004 were filled in using the 2003, 2000, 1997, and 2003 editions, respectively. In this way, we keep consistent validation methods even when official post-digital texts are missing.

To streamline our analysis and validations, we group titles into four legal categories. There are 50 titles when the U.S. Code was first enacted in 1926, each corresponding to a broad domain of federal statutory law—such as taxation, public lands, criminal justice, or national defense. And thus, we use these titles as the primary organizing units in this paper. We find some titles were densely populated with detailed provisions, while others were largely empty, marked as reserved or sparsely populated in anticipation of future legislative development. We group these titles into four major functional domains based on their perceived function: (1) Government structure: Titles 1, 2, 3, 4, 5, 13, 36, and 44; (2) National defense and foreign affairs: Titles 10, 14, 22, 32, 37, 38, 48, and 50; (3) The economy: Titles 11, 12, 15, 19, 23, 26, 30, 31, 33, 39, 40, 41, 43, 45, 46, 47, and 49; (4) Society and civil life: Titles 7, 8, 9, 16, 17, 18, 20, 21, 24, 25, 27, 28, 29, and 42. However, the full set of titles is available such that future researchers can categorize themselves.

Finally, our dataset includes title-year files. This data structure is needed because titles were subdivided, renumbered, merged, or repealed, reflecting changing policy priorities, administrative needs, and the increasing complexity of federal governance. For example, Title 6 was originally designated for "Surety Bonds" but was repurposed as "Domestic Security" in 2002 following the September 11, 2001 attacks. Title 34 was initially titled "Navy" and merged into Title 10 ("Armed Forces") in 1956. It remained inactive until 2017, when it was reinstated under a new designation: "Crime Control and Law Enforcement." Title 51 was introduced in 2010 to consolidate statutes related to "National and Commercial Space Programs." Title 52 was created in 2014 to organize laws pertaining to "Voting and Elections." Title 54 created in 2014 for "National Park Service and Related Programs." Additionally, Title 53 is currently reserved for "Small Business," but the proposed legislation composing this title has not yet been passed by Congress [20]. Accordingly, we track the growth and evolution of 48 continuous titles from 1934 to the present. References to "uncleaned text" below refer to the generative AI post-processed versions of the U.S. Code data, and "cleaned text" refers to the uncleaned text with duplicate spacing removed and footer and header patterns removed prior to 1994. Although the full set of title-year files is released, our validation method excludes those discontinued titles.

## Data Set 1: Contents of U.S. Code

The U.S. Code is, fundamentally, a large and evolving text corpus—composed of words, vocabulary, and structure—that can be analyzed using tools from computational linguistics and information theory. Words were extracted using the `word tokenize` function from the `NLTK` Python library, and all words were lowercased and alphanumeric. Measures such as total word count, vocabulary size, and word entropy offer insight into the growth, complexity, and lexical redundancy of the Code over time. The total amount of text serves as a proxy for the informational volume of the legal corpus. The number of unique words captures the degree of lexical novelty, while normalized Shannon entropy indicates the distributional diversity of language, concentration on certain legal topics, use across editions. In addition, analyzing these statistical patterns in our OCR-processed text—and comparing them with those observed in other large natural language corpora—provides a useful check on the plausibility and quality of our data, while also testing whether legal text exhibits scaling regularities similar to those found in general-purpose language. To reduce OCR-related noise, all statistics were computed using only words that appear at least twice in the full text of the U.S. Code for each year. Accordingly, Data Set 1 includes the following features for each title $i$, at a given time $t$. We also report character and word counts for the uncleaned OCR text prior to 1994 for comparative purposes.

- $N_i(t)$: total length of the cleaned text (in characters).
- $W_i(t)$: total word count of the cleaned text.
- $V_i(t)$: number of unique words (vocabulary) in the cleaned text.
- $S_i(t)$: Shannon entropy of the word-frequency distribution, defined as $S_i(t) = -\sum_{k=1}^{V_i(t)} p_k \log_2 p_k$, where $p_k = n_k(t)/W_i(t)$, and $n_k(t)$ is the frequency of word $k$.
- $\hat{S}_i(t)$: normalized Shannon entropy, $\hat{S}_i(t) = S_i(t)/\log_2(V_i(t))$ such that it is bounded between 0 and 1.
- $S(t)$: Shannon entropy of the word-frequency distribution over all titles.
- $\hat{S}(t)$: normalized Shannon entropy of the word-frequency distribution over all titles.

Fig. 3 presents the growth of each of the 48 titles in our dataset over time, measured as $W_i(t)$, the number of cleaned words in title $i$ in year $t$. The inset displays the aggregate word count across all titles, $\sum_{i=1} W_i(t)$, on a logarithmic scale. Among these, we select the four largest titles by $W_i(t = 2023)$—Title 42 (The Public Health and Welfare), Title 26 (Internal Revenue Code), Title 10 (Armed Forces), and Title 16 (Conservation)—in Fig. 4, with insets showing their growth on a logarithmic scale. Each figure includes a vertical dashed line marking the year 1994, the transition from OCR-based scans to digitalized datasets. The smooth transition between these two data sources—both at the level of individual titles and in aggregate—suggests the consistency and reliability of our OCR cleaning process.

Beyond corpus growth, we also examine the internal structure of language use by measuring normalized Shannon entropy $\hat{S}_i(t)$, shown in Fig. 5. Shannon entropy has been proposed as a distinguishing metric for legal texts, differentiating them from other genres, such as novels and plays, as well as from legal documents written in other languages [21]. We observe a clear decline in $\hat{S}(t)$ for the full U.S. Code over time, as well as the three largest titles. Again, we provide a vertical mark for the two different data sources, showing the persistence and smoothness of these trends, reinforcing the reliability of our reconstructed corpus for large-scale analysis.

Next, we measure how vocabulary growth $V_i(t)$ in each title $i$ by measuring the number of unique words, as shown in the insets of Fig. 5. Together with the total word count $W(t)$, we find our OCR-processed text exhibits the well-known empirical signature of Heaps' law: $V_i \sim W_i^{\beta_i}$ [22, 23]. Fig. 6(A) and (B) show the relative growth for Title 19 (Customs Duties) and Title 4 (Flag and Seal, Seat of Government, and the States), the slowest growth ($\beta = 0.25$) and the fastest growth ($\beta = 0.71$), respectively. The distribution of scaling exponents $\beta$ across all titles is presented in Fig. 6(C). The bulk of these values cluster around $0.5$, consistent with empirical findings in other large-scale language corpora such as newspapers and web pages [24, 25]. Altogether, the lexical scale pattern and their exponents suggest that our OCR-processed text of U.S. Code aligns with general statistical properties of natural language.

The variation in $\beta$ suggests substantial variation in the relative growth in vocabulary across legal domains, leaving room for future empirical studies. We provide additional statistics, combining with $W_i$. In Fig. 6(C), titles are color-coded by their total word count $W_i$, from blue (small) to red (large), showing a negative correlation between corpus size and $\beta$: smaller titles tend to generate new vocabulary at a higher rate. This suggests that different areas of law impose distinct demands on lexical diversity, a pattern that sets legal language apart from more colloquial genres like fiction or journalism.

We additionally estimate and report the well-known Zipf law exponent, which characterizes corpus growth beyond mere quantitative increases in word count. Unlike simple measures of corpus size, this exponent captures more nuanced aspects of linguistic development—specifically, the degree of lexical diversity and structural complexity within the text. Given that legal texts such as the U.S. Code are composed of highly specialized terminology, their word frequency distributions are often skewed by a small number of extremely frequent terms. To address this, we adopt the generalized Zipf-Mandelbrot law, expressed as $F \sim 1/(r+b)^a$, where $F$ is the frequency, $r$ is the rank, $b$ is the shift parameter, and $a$ is the scaling exponent. We estimate $a$ for each title and year of the U.S. Code to examine how language structure evolves across the legal corpus. Specifically, we fit the Zipf–Mandelbrot function using non-linear least squares optimization (`scipy.optimize.curve_fit`), minimizing the residual sum of squares between the observed word frequencies and the fitted curve. All code used for estimating the exponents and reproducing the results is available on GitHub. As shown in Fig. 7, the rank-frequency distributions of all titles, Title 19, and Title 4 closely follow the Zipf-Mandelbrot law, with high goodness-of-fit. This pattern is not limited to these examples—most Titles exhibit similar adherence to the Zipf-Mandelbrot distribution. Moreover, we find a negative correlation between the Zipf-Mandelbrot exponent ($a$) and the Heap's law exponent ($\beta$), as illustrated in the inset. This relationship is consistent with the theoretical expectation, indicating that corpora with greater lexical diversity (higher $\beta$) tend to exhibit flatter word frequency distributions (lower $a$), reflecting reduced dominance of the most frequent terms.

We further trace the Zipf-Mandelbrot exponent $a$ over time, from 1934 to 2023, across all titles. Notably, the transition around 1994—marking the shift from OCR-processed text and web-crawled text—shows no structural break, supporting the reliability and continuity of our dataset. Similar smooth temporal patterns are observed in individual Titles (see Fig. S1.7), although some titles exhibit sharp spikes attributable to major legal expansions. These anomalies are further discussed in the Supplementary Information (see Fig. S1.8.)

Additional analyses using Data Set 1 are included in the Supplementary Information: word count growth across all 48 titles (Fig.S1.1); changes in unique vocabulary size (Fig.S1.2); vari-ation in normalized entropy (Fig.S1.3); Heaps' law estimates for each title (Fig.S1.4); and Zipf-Mandelbrot's law for each title in 1934 and 2023 (Fig.S1.5, S1.6)

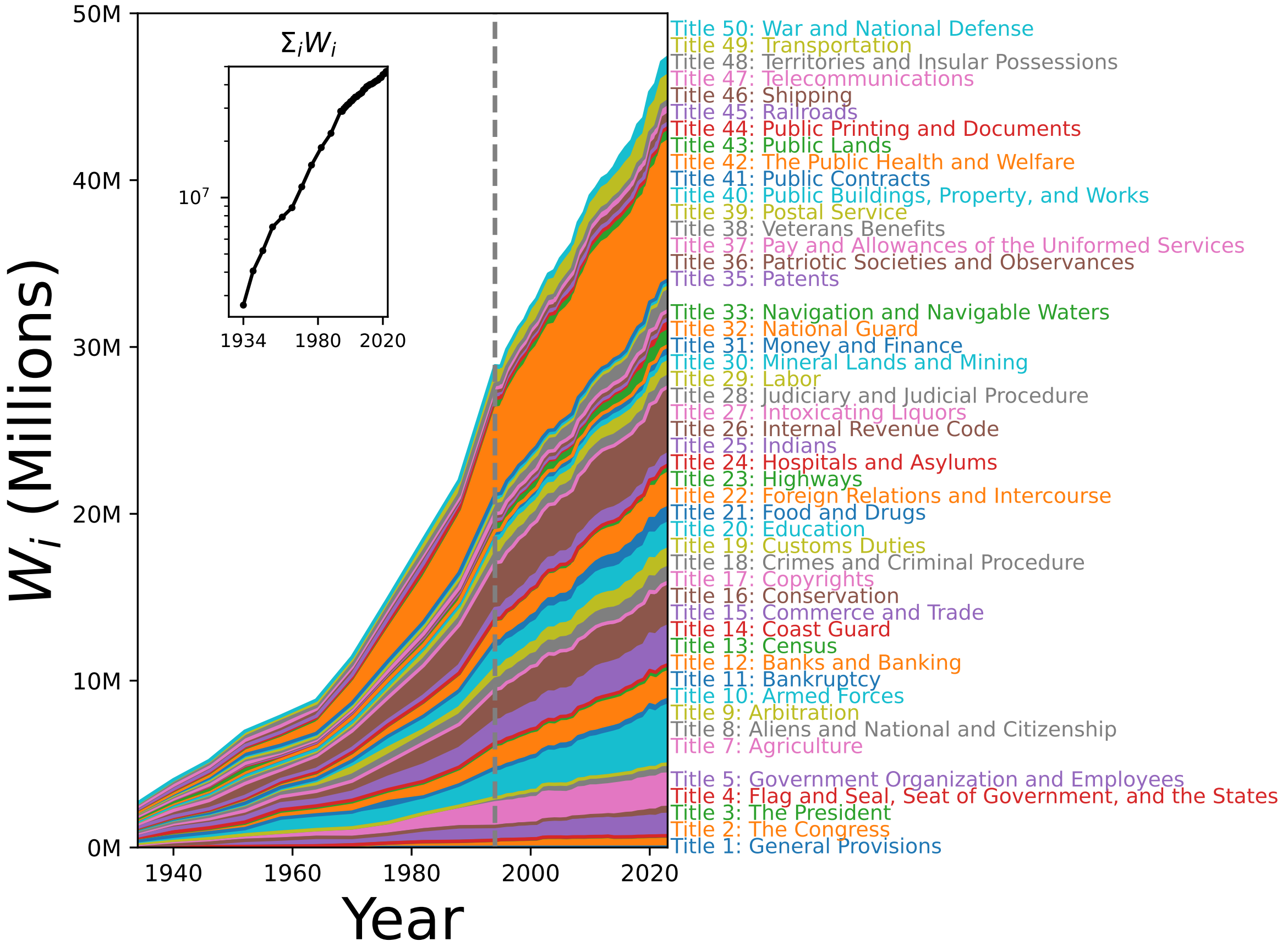

**Figure 3: The total content of the U.S. Code has steadily increased over time.** Stacked area chart showing the number of cleaned words in each title $i$, denoted $W_i$, from Title 1 (General Provisions) to Title 50 (War and National Defense). The inset displays the total word count across all titles, $\sum_{i=1}^{50} W_i(t)$, excluding the discontinued titles 6 and 34, on a logarithmic scale. A vertical dashed line marks the year 1994, the transition from OCR-based data to web-based digital texts.

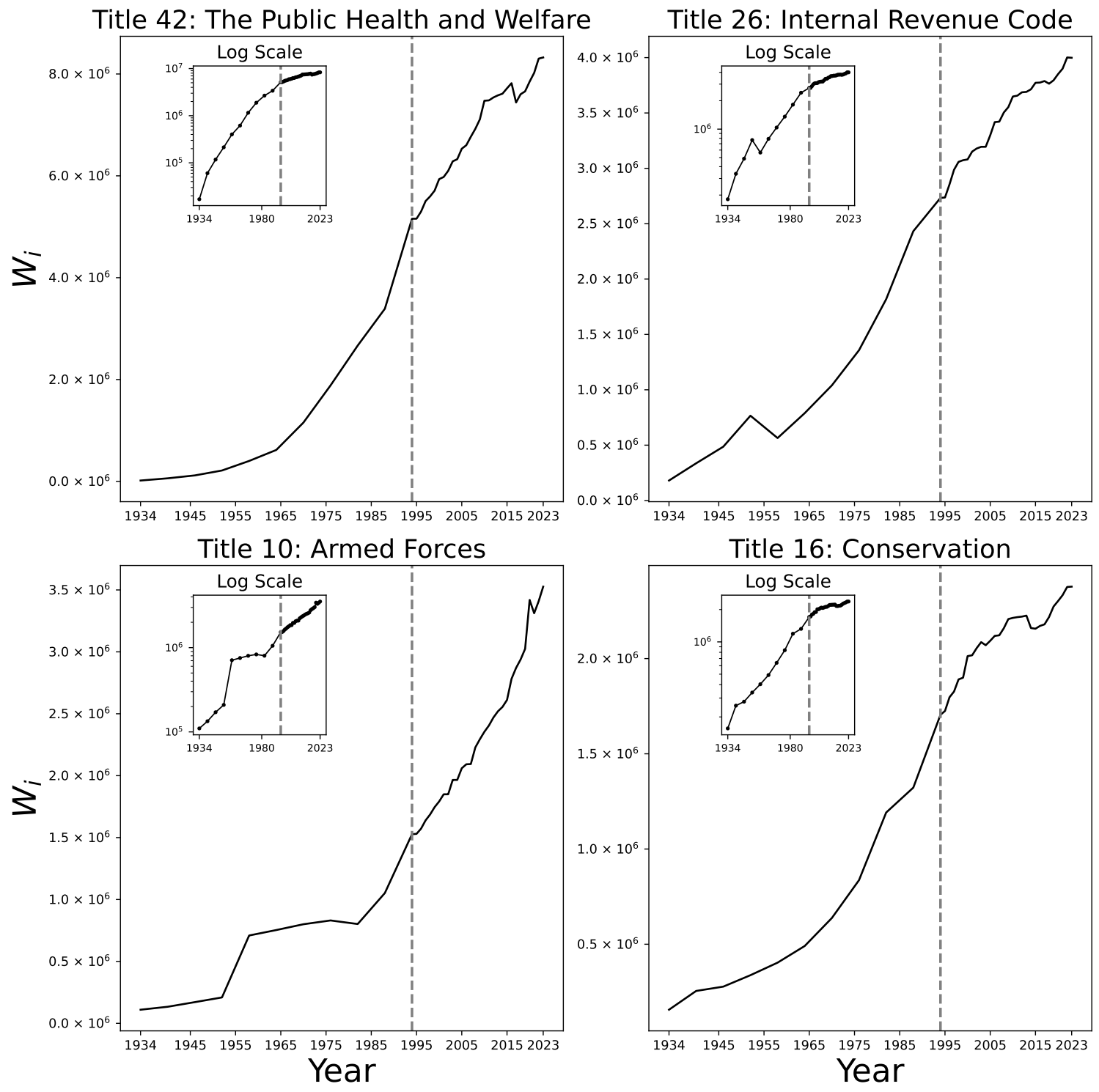

**Figure 4: The titles with the highest word counts in the U.S. Code.** Among all titles, Title 42 (The Public Health and Welfare), Title 26 (Internal Revenue Code), Title 10 (Armed Forces), and Title 16 (Conservation) have the highest word count. The $y$-axis represents the number of cleaned words and the $x$-axis denotes the year. A vertical dashed line marks the year 1994, indicating the transition between OCR data and web data. In each plot, an inset shows a log-linear plot of the growth of each title.

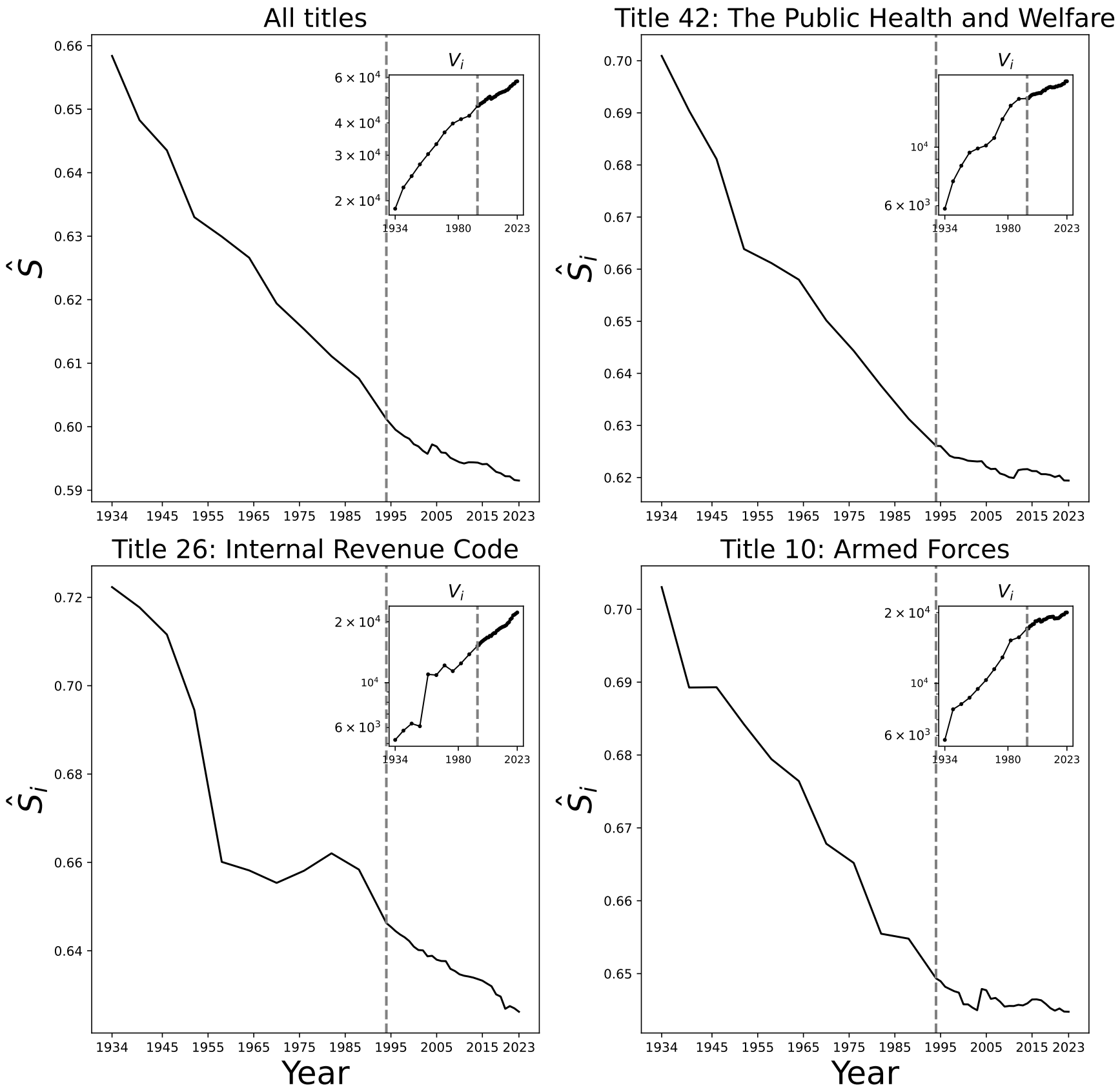

**Figure 5: Declining normalized Shannon entropy of the U.S. Code over time.** The normalized entropy of the U.S. Code has been decreasing over time, indicating standardizing legal contents. A vertical dashed line marks the year 1994, indicating the transition between OCR data and web data. Note that the data sets are smoothly connected. In each plot, an inset shows a log-linear plot of the unique word count growth of each title.

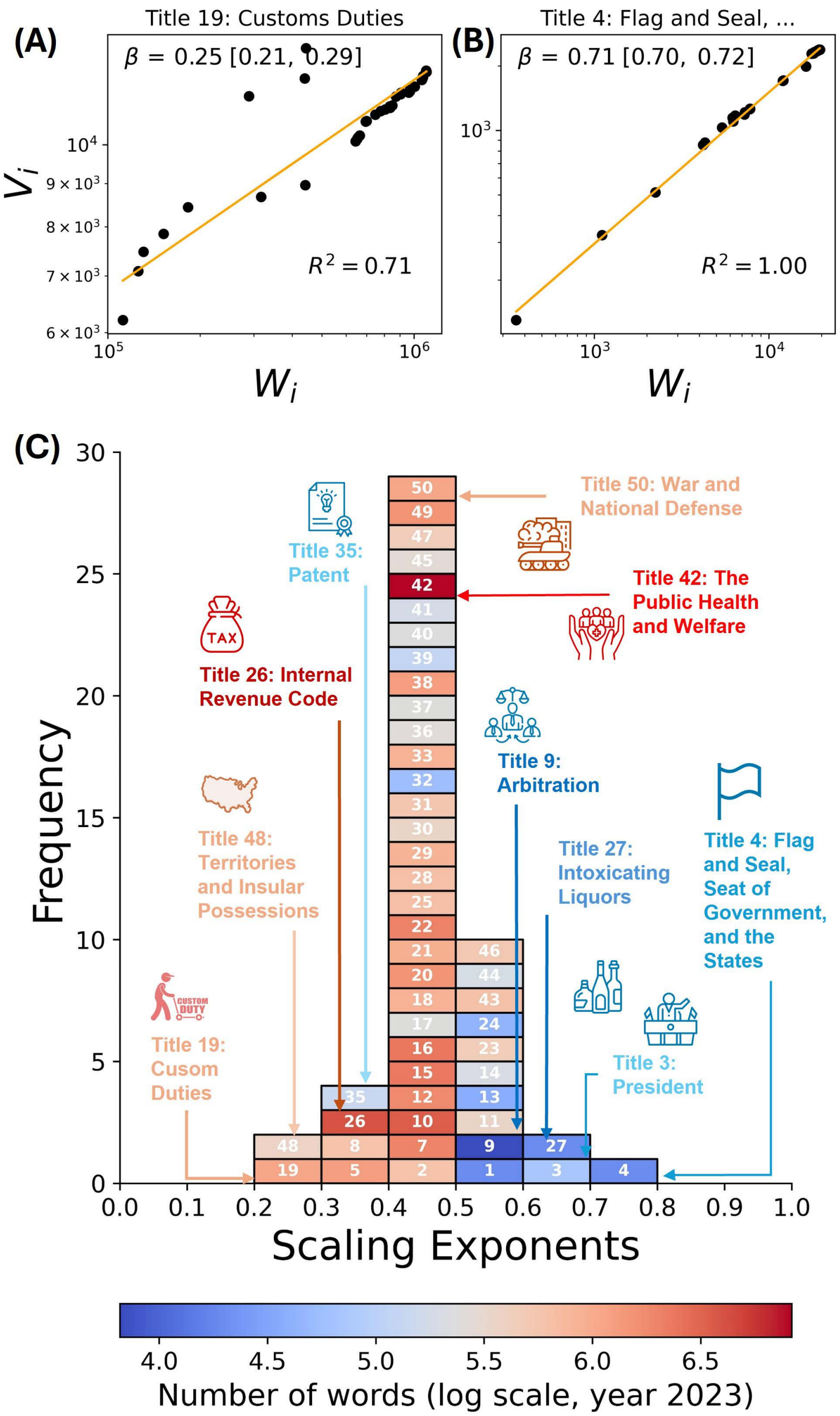

**Figure 6: Heaps' law in the U.S. Code.** The contents of the U.S. Code follow Heap's law, demonstrating a power-law relationship between the total number of cleaned words ($x$-axis) and the number of unique cleaned words ($y$-axis). Each scatter point represents yearly data for a specific title. The average $\beta$ value is 0.47, with (A) the lowest $\beta = 0.21$ for Title 19 and (B) the highest $\beta = 0.71$ for Title 4. (C) The scaling exponent of 48 titles has the highest frequency between 0.4 to 0.5.

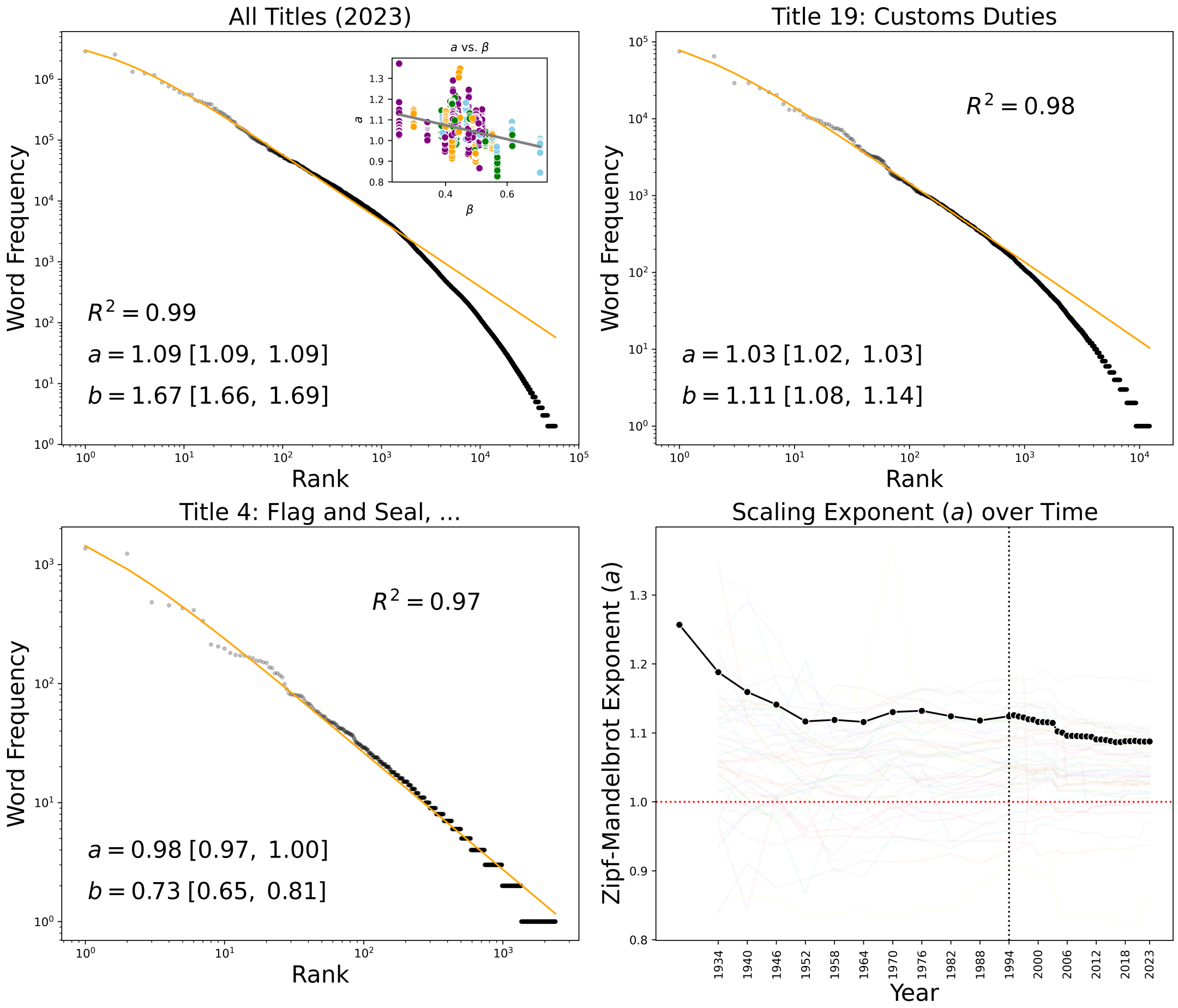

**Figure 7: Zipf-Mandelbrot's law in the U.S. Code.** Rank-frequency distributions of words for all titles in the 2023 U.S. Code, Title 19 (Customs Duties), and Title 4 (Flag and Seal, Seat of Government, and the States). Each plot is fitted with the Zipf-Mandelbrot model, with parameters $a$ (scaling exponent), $b$ (shift parameter), and corresponding $R^2$ values indicating fit quality. Temporal trend of the scaling exponent $a$ from 1934 to 2023 for each titles and for all titles (black). A vertical dashed line marks the year 1994, indicating the transition between OCR data and web data. Note that the data sets are smoothly connected. In the first plot, an inset shows a relationship between Scaling exponent from Heap's law and Zipf-Mandelbrot's law.

## Data Set 2: Hierarchical Structure

The complexity of the U.S. Code arises from multiple dimensions. In addition to word-level statistics such as frequency and distribution, its hierarchical organization infers internal logic through which each complex legal domain is unpacked and specified. As such, each title comprises multiple layers of nested units, such as *Subtitles*, *Divisions*, *Parts*, *Subparts*, *Chapters*, *Subchapters*, *Sections*, *Subsections* (lowercase letters), *Items* (numbers), *Subitems* (uppercase letters), where finer divisions are marked by Roman numerals in both cases. Fig. 1 Data set 2 illustrates this hierarchy as a tree structure, from a common root node $R$ (Title $i$) to each node $v$ with depth $d$. Once mapped onto such a tree, we quantify each title's structural metrics and their time evolutions in a consistent way, such as the maximum and average breadth and depth per title: $\max(d)$, $\langle D \rangle$, $\max(b)$ and $\langle B \rangle$.

Like many semi-structured textual corpora, however, the U.S. Code presents nontrivial chal-lenges for parsing due to inconsistencies in formatting and deviations from the standard hierarchical template. For example, while all titles in the U.S. Code contain at least a *Chapter* and *Section* structure, not all of them include the full set of hierarchical units. Several titles omit one or more intermediate levels—such as *Subtitles*, *Divisions*, *Parts*, or *Subparts*—between the *Title* and *Chapter*, leading to structural irregularities that must be handled carefully in extraction and representation. We manually went through the documents and find the structural conventions of each specific title, which requires separate parsing rules within the document—high-level and low-level structure.

As the U.S. Code evolves, new hiearchical layers are sometimes introduced. For example, Title 41 (Public Contracts) adopted additional hierarchical layer such as *Subtitles* and *Divisions* after 2010. In other cases, the order of structural units differs from the expected norm. In other cases, the order of structural units is reversed. For example, Title 16 (Conservation), Title 26 (Internal Revenue Code), and Title 42 (The Public Health and Welfare) place *Parts* under *Subchapters*, contrary to the expected nesting order. These dynamic characteristics of the U.S. Code necessitate adaptive parsing strategies capable of handling structural variation.

In summary, we customize the parsing logic to reflect the unique structural features of each title. Our approach makes use of hierarchical cues—such as indentation levels, alphanumeric labels (e.g., "(a)", "(1)", "A."), and recurring textual patterns (e.g., "Section", "Subchapter")—to identify the nested organization of legal provisions. In the following sections, we describe a modular, multi-step pipeline that incorporates these cues through rule-based pattern matching, indentation-aware logic, and graph-based post-processing. This pipeline allows us to accurately reconstruct the legal hierarchy, from high-level units like *Subtitles*, *Parts*, *Chapters*, and *Subchapters*, down to more detailed components such as *Sections*, *Subsections*, *Items*, and their *subcomponents*.

**First step: high-level hierarchy**

We begin with a simple, regular-expression-based rule for the upper layers of the legal hierarchy units, such as *Subtitles*, *Divisions*, *Parts*, *Subparts*, and *Chapters*. After this initial pass, a reorganization step nests *Chapters* under the most recent preceding *Subpart*, where appropriate.

We then parse the *Chapter* $\rightarrow$ *Subchapter* $\rightarrow$ *Section* hierarchy, placing each *Section* either under its corresponding *Subchapter* or directly under the *Chapter* when no *Subchapter* exists. These components are merged to construct a unified *high-level hierarchy*, connecting the *Title-to-Chapter* and *Chapter-to-Section* hierarchies. And 10 titles (Title 2, Title 6, Title 8, Title 16, Title 20, Title 26, Title 34, Title 42, Title 47, Title 50) follow the hierarchical order in which *Parts* appear after *Subchapters*. For these titles, we apply custom parsing logic that respects this ordering by placing *Parts* after *Subchapters*.

**Second step: low-level hierarchy**

The low-level structure such as *Subsections*, *Items*, *Subitems*, and further down levels requires an indentation-aware parser because these elements are assigned structural classes based on indentation depth and textual patterns. Once again, due to textual irregularity, there are cases where this approach—relying on regular expressions and indentation—fails to distinguish between similarly denoted units, such as a lowercase-letter *Subsection* (e.g., (i)) and a Roman numeral-style *Subitem* (e.g., (i)).

We therefore develop a post-processing routine that includes: (i) correcting misclassified Roman numerals based on local contextual cues, and (ii) eliminating adjacent duplicate entries by retaining the one with higher structural priority. The cleaned elements are then transformed into a *low-level hierarchy* using a recursive parent-seeking logic, where each node is linked to the most recent valid upper-level node—such as nesting a *Subitem* under the latest *Item*, or a *Subsection* under a corresponding *Section*—even when intermediate layers are skipped in the text.

**Final Step: Tree structure construction**

We merge the *high-level hierarchy* and the *low-level hierarchy* into a single hierarchical graph for each title and year. This is accomplished by identifying shared *Section* nodes and incorporating their descendant subtrees from *low-level hierarchy* into the *high-level hierarchy*. The resulting merged tree captures both the high-level legal structure (e.g., *Titles*, *Subtitles*, *Parts*, *Subparts*, *Chapters*, *Subchapters*) and the low-level hierarchy (e.g., *Sections*, *Subsections*, *Items*, and more granular elements). The tree for each title can reach a maximum depth of 11, following the path: *Title* → *Subtitle* → *Part* → *Subpart* → *Chapter* → *Subchapter* → *Section* → *Subsection* (lowercase letter) → *Item* (number) → *Subitem* (uppercase letter) → Roman numeral (lowercase) → Roman numeral (uppercase).

For final validation, the merged graph is constructed to be compatible with Python's `networkx` library such that the tree structure has a single-root and acyclic properties. All intermediate outputs—including nested data structures, structural component labels with corresponding text, and tree representations—are saved and made available to support reproducibility and downstream analysis.

The final outputs are released in two formats for each title: (1) a `networkx` tree object and (2) CSV files containing detailed node-level information. These formats support flexible use for visualization, replication, and further analysis. For each title, we provide three variants of the `networkx` tree object:

- *High-level hierarchy*: Represents the hierarchical structure from the *Title* level down to *Sections*.
- *Low-level hierarchy*: Captures the fine-grained hierarchy from each *Section* to its most granular components (e.g., *Subsections*, *Items*, *Subitems*, etc.)
- *Entire tree*: A merged hierarchical structure that integrates the *high-level hierarchy* and *low-level hierarchy* for each title.

Each of the three structural formats offers its own advantages. The *high-level hierarchy* provides a clear overview of the legal structure, making it suitable for visual comparisons and large-scale structural analyses. The *low-level hierarchy* supports fine-grained analysis of legal content by capturing the detailed organization of subsections and clauses. The entire tree merges both levels as in Fig 9, allowing multi-level network analyses and the study of hierarchical transitions from general titles to specific legal provisions. By separating the structures, users can flexibly choose the appropriate level of granularity depending on their analytical objectives, while optimizing computational efficiency and enabling scalable applications across diverse research contexts.

### Shape metrics for Hierarchical tree

Hierarchical tree structures represent this organization, as illustrated in Fig. 1. We compute the following structural features for each title of the *entire tree* data (third structural format): (1) the depth of the tree $d$ (i.e., the number of hierarchical levels); (2) the breadth $b_l$ at each level $l$ (i.e., the number of nodes per level); (3) the total number of leaf nodes; and (4) the branching factor, which quantifies the degree of structural expansion at each node within the hierarchy. Data Set 2 includes these features for each title $i$.

- $T_i(N_i, L_i)$: Entire tree network rooted at a title node, with node set $N_i$ and link set $L_i$.
- $|N_i|$: Number of nodes.
- $d_n$: Depth, defined as the length of the path from the root (title) to node $n$.
- Max Depth: Maximum depth, defined as $\max_n d_n$.
- $\langle D_i \rangle$: Average depth, calculated as $\sum_{n \in N_i} d_n / |N_i|$.
- $b_k$: Breadth, the number of nodes at depth $k$.
- Max Breadth: Maximum breadth of the tree, defined as $\max_k b_k$.
- $\langle B_i \rangle$: Average breadth, calculated as $\sum_{k=0}^{\text{Max Depth}} b_k / (\text{Max Depth} + 1)$.
- $\text{Leaf}_i$: Number of leaf nodes (nodes with zero out-degree).
- $\langle \phi_i \rangle$: Average branching factor, defined as the average number of outgoing links (out-degree) for non-leaf nodes.
- $\text{Var}(\phi_i)$: Variance of branching factor.

Fig. 8 shows the temporal evolution of the structural statistics of the tree representations (the mean depth and breadth) for each title over time. For simplicity, we group titles into four thematic domains of governance as introduced in Section *Data Curation for the U.S. Code*: government structure, national defense, economy, and society. Most titles have evolved primarily in terms of breadth expansion, suggesting that the U.S. Code has been increasingly incorporating a wider range of legal considerations, regardless of title or domain.

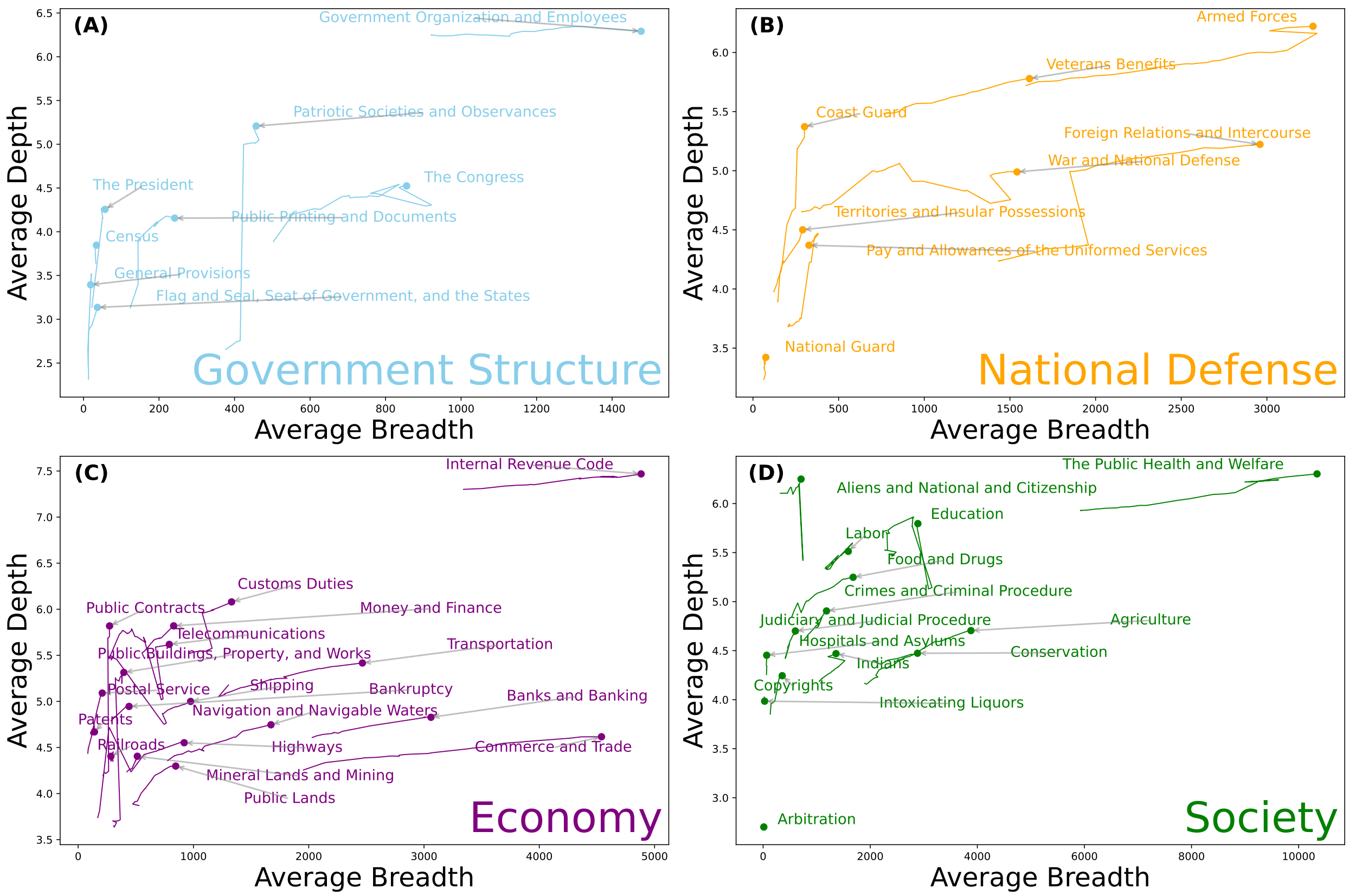

**Figure 8: Structural Evolution of the U.S. Code from 1994 to 2023.** The hierarchical structure of the U.S.Code shows a consistent trend toward breadth expansion across most titles. The trend suggests that the Code has increasingly incorporated a wider range of legal considerations over time, from topical diversification to specifications through hierarchical layering, across titles or domains. Titles related to economic and societal functions especially exhibit concurrent growth in both depth and breadth, while those associated with government structure and national defense tend to grow along a single dimension. For an alternative view using a linear-log scale, see Fig. S1.9 in the SI.

In each tree, node labels are constructed using the structural class name along with its identifier (e.g., `Chapter_3`, `Section_1`). However, because structural class names and identifiers may repeat within the same title—such as `item_2` appearing under both `subsection_a` and `subsection_b`—each node is uniquely identified by its full hierarchical path from the corresponding *Section* node. For instance, a node may be labeled as `section_204_subsection_b_item_2`. This naming convention guarantees path uniqueness, a fundamental property of tree-based structures, and eliminates ambiguity arising from repeated combinations of class names and identifiers, which are especially common below the section level.

Detailed node-level information is also provided in the form of a CSV file. This file includes the node label used in the Tree object and the corresponding legal sentence associated with each node (e.g., `Title 1` - "General Provisions"; `section_2205_item_4` - "(4) the return of spoils of war ...").

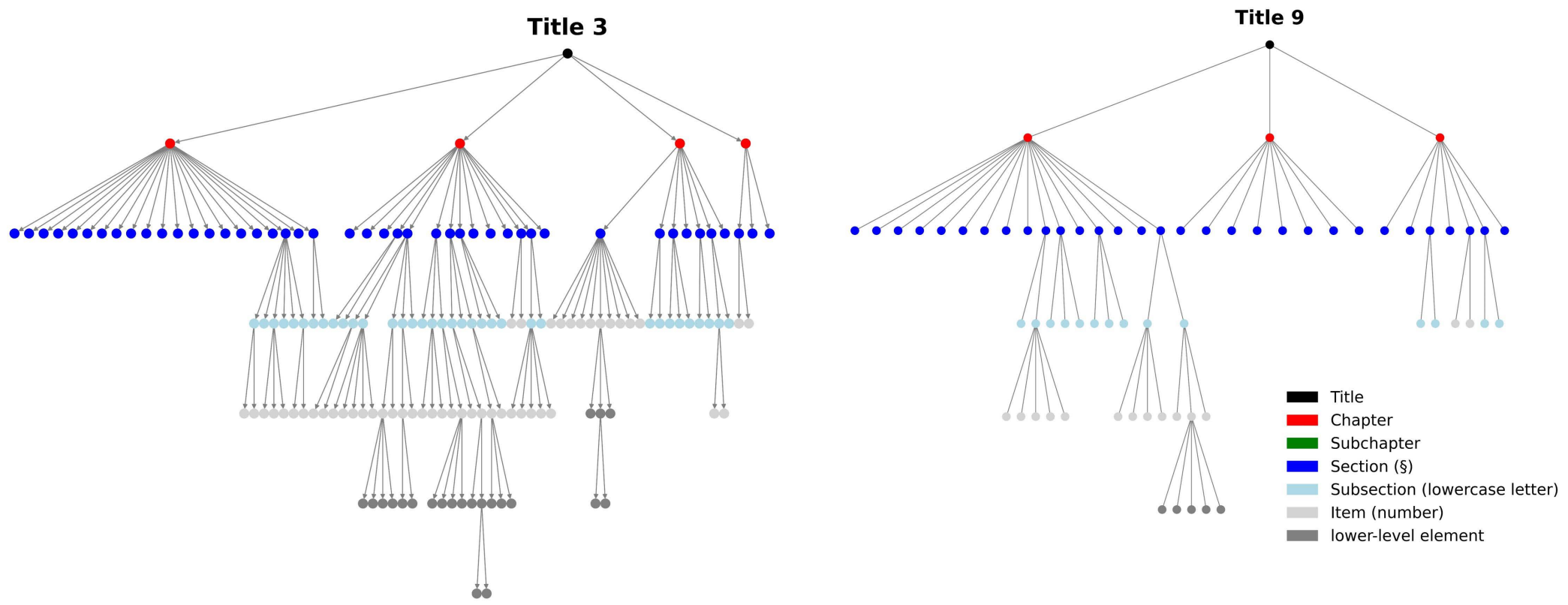

**Figure 9: Tree structure of titles.** (A) Tree structure of Title 3 (The President) for the year 1994. In 1994, Title 3 had an average depth of 3.14, a maximum depth of 6, an average breadth of 22.14, a maximum breadth of 49, and 114 leaf nodes. (B) Tree structure of Title 9 (Arbitration) for the year 1994. In 1994, Title 9 had an average depth of 2.74, a maximum depth of 5, an average breadth of 11.33, a maximum breadth of 31, and 53 leaf nodes.

Using the provided tree objects and node information, researchers can reproduce and analyze hierarchical structures such as those illustrated in Fig. 9. These resources support custom visualizations and structural analyses across titles and years. Note that we do not provide tree structures prior to 1994 due to potential errors in the OCR-extracted text from historical U.S. Code volumes. The generative AI methods used to reconstruct the text did not preserve indentation or spacing information, making it infeasible to reliably infer the hierarchical structure of titles for earlier years. Nevertheless, the full reconstructed text for these earlier editions (title-year files) is included in the dataset to support future improvements in structure recovery.

## Data Set 3: Cross-Reference Relationship

No legal domain operates in isolation. Each title in the U.S. Code frequently references other titles to incorporate relevant legal information across domains. Fig. 1 (Data 3) illustrates how these cross-references are parsed using a sentence-level pattern-matching procedure. For each year and for all 50 titles, we retrieve the full cleaned text and segment it into individual sentences using the `NLTK` library's pretrained tokenizer, which reliably identifies sentence boundaries. We then use a regular expression to search for occurrences of the phrase pattern `"title ?"`, where `?` denotes any one- or two-digit integer. All sentences containing such patterns are extracted and stored as entries in a cross-reference edge list.

For each sentence that includes a reference to a specific title (e.g., `title 5`), we count the number of times that title is mentioned. Our final data structure is represented as a directed, weighted network of cross-title references for each year, where each edge from `title i` to `title j` represents the number of times `title j` is explicitly cited within the text of `title i`, providing a consistent and interpretable baseline for quantifying the structural and semantic connectivity within the *U.S. Code* over time. The final Data Set 3 contains three parts:

- *Cross-Reference*: Cross-referencing relationships between titles in the Code from 1934 to the present;

- *Edge list*: An edge list of these references, including the citing text;
- *Backbone structure*: The backbone structures of the resulting cross-reference networks across years.

The first and second parts of Data Set 3 are structured in a table format with three columns: the citing title, the cited title, and the number of citations between them. As such, the dataset contains the full records of parsed cross-referencing relationships between titles with detailed cross-referencing edge information in the *U.S. Code*. We provide additional metadata such as the names of the citing and cited titles, the cleaned word counts of each title, and the citation ratios—defined as the number of citations normalized by the word counts of the citing and cited titles, four major functional domains of each citing and cited title. From these data, one can construct an adjacency matrix of cross-references for each year of the U.S. Code.

In addition to these detailed datasets, we provide a backbone network that only includes statistically or semantically significant links for researchers' convenience. For this backbone structure, we use a well-established *disparity filter* to filter out cross-references, likely arisen by chance, given a null model in which weights are randomly distributed [26, 27, 28]. We applied python package [29] using [26]. For each year, we calculate optimal $\alpha$, the significance threshold for retaining a link, to yield a meaningful structure. More specifically, we compare the outcomes of two edge-filtering methods [26]: a global threshold-based backbone (GTB) filter and the disparity filter backbone (DB). For each year, we construct citation networks with weighted edges representing citation strength and apply GTB filters using empirical weight quantiles (50th–99th percentiles). We then apply the DB filter across a range of significance levels ($\alpha$) and compute the overlap fraction between the resulting edge sets and the GTB sets. The optimal $\alpha$ gives the smallest value at which the DB-filtered network fully recovers the set of links corresponding to the 99th-percentile GTB network. This approach provides a consistent, data-driven method for calibrating the disparity filter based on the empirical distribution of edge weights. The yearly results for optimal $\alpha$ values are presented in Fig. S1.10. All of these processes can be done with the fully detailed dataset that we released.

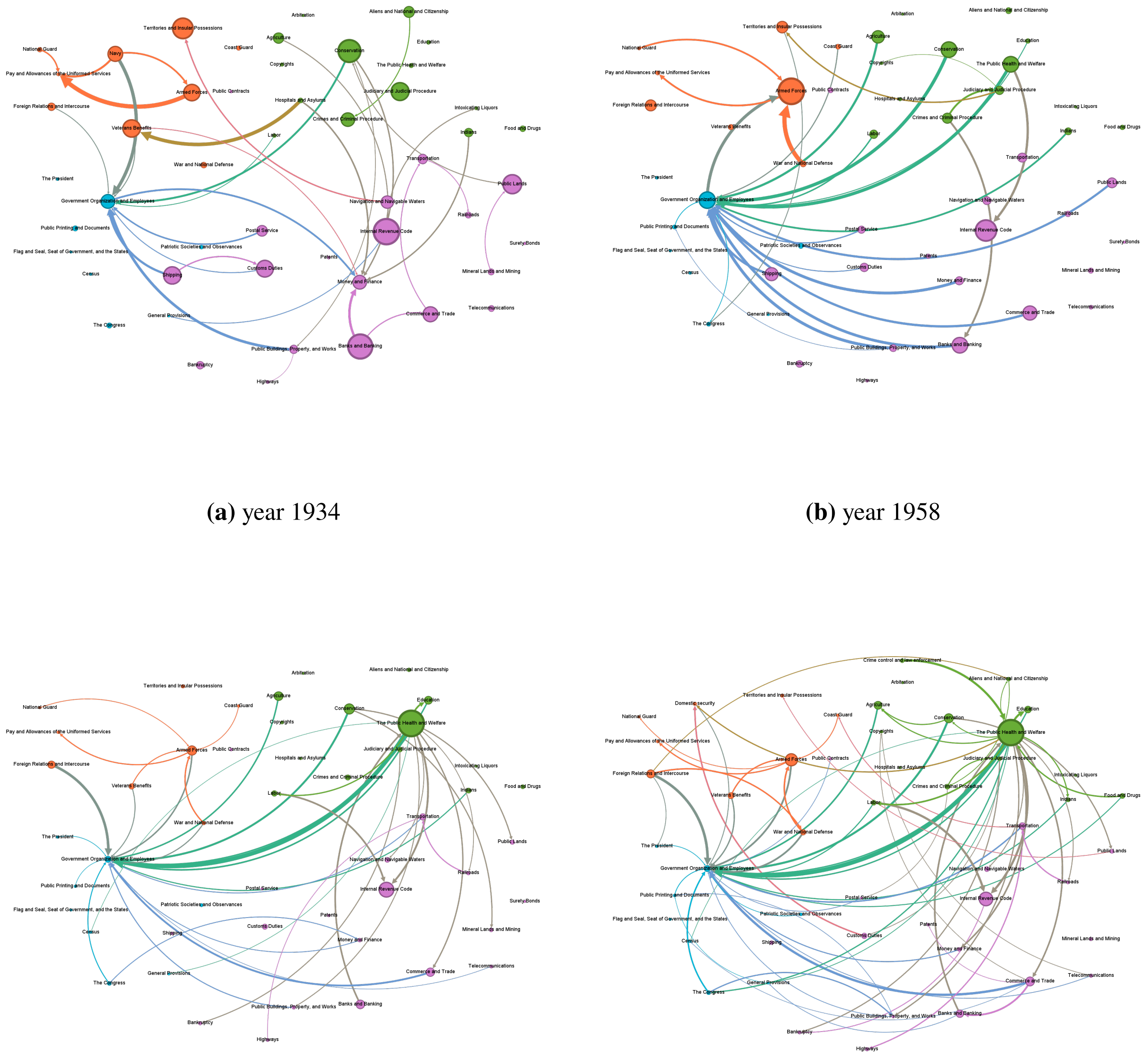

**(a)** year 1934

**(b)** year 1958

**(c)** year 1994

**(d)** year 2023

**Figure 10: Backbone structure of the U.S. Code's title cross-reference network in 1934, 1958, 1994, 2023.** Nodes and directed links represent titles, and citations in the backbone network (see Methods and SI). Node size and link weight are proportional to word counts ($W_i$) and cross-referencing between two titles, respectively.

Figs. 10 shows the backbone structures of the *U.S. Code* in 1934, 1958, 1994, 2023 (see the network of the entire year in SI Fig. S1.11, Fig. S1.12, Fig. S1.13, Fig. S1.14). The network becomes denser and the central titles shift. For example, "Government Organization and Employees" gradually increases (Fig. 10(b)) and "The Public Health and Welfare" (Fig. 10(c)). In particular, once insignificant "The Public Health and Welfare" has become the largest and one of the most well-connected titles by 1994 and especially in 2023 (Fig. 10(d)). The structure seems to be stabilized by 1994 and instead the inter-dependencies become denser until 2023 (Fig. 10(d)).

The third part of Data Set 3 contains the necessary information to reconstruct directed graphs representing the backbone structures of the cross-reference networks for each year of the *U.S. Code* from 1934 to the present. This information is provided in list format with the same columns and with a cross-reference part. Users who wish to generate sparser or denser backbone networks—corresponding to different values of the parameter $\alpha$—can use the full set of cross-references provided in the first part of Data Set 3, in combination with the backbone extraction algorithms cited above, to construct networks at alternative levels of resolution.

Future analyses that could be performed using Data Set 3 include: modeling the U.S. Code as a legal hypergraph [30], rather than as the pairwise graphs or networks provided here; performing network motif analysis [31], a technique commonly used in the study of gene-regulatory networks; and applying community detection algorithms to identify clusters of closely related titles [32].

# Data Records

The dataset is freely available at this Figshare link.

## Dataset 1: Contents of U.S. Code

Table 1 summarizes the data for a single U.S. Code title in a given year. The table distinguishes between raw and cleaned text fields: raw fields capture the statutory text as originally digitized, while cleaned fields have been processed to remove OCR artifacts, headers, and extraneous content for greater accuracy. Entropy values, calculated from the cleaned text, quantify the linguistic diversity and complexity of each title. Scaling exponents are derived from statistical fits to the vocabulary growth (Heaps' law) and rank-frequency (Zipf-Mandelbrot law) distributions, offering insight into the underlying linguistic patterns. Finally, each title is assigned to a functional group, enabling comparative analyses across different domains of the U.S. Code.

## Dataset 2: Hierarchical Structure

Dataset 2 represents the internal legal hierarchy of each U.S. Code title as a tree structure, capturing the nested organization from Title to Chapters, Sections, Subsections, and beyond. For every year and title, we construct three types of tree networks—high-level tree, low-level tree, and entire tree—and provide them in GML format. This approach enables detailed analysis of the hierarchical structure at multiple levels of granularity.

The `Tree_stats.csv` file contains summary statistics in *Shape metrics for Hierarchical tree* characterizing the hierarchical structure of each U.S. Code title for every year. Each row represents a unique title-year combination. Table 2 provides a comprehensive overview of the structural metrics computed for each tree, including measures of depth, breadth, branching, and overall complexity.

The `node_meta.csv` file provides metadata for each node at or above the section level in the hierarchical representation of the U.S. Code. Each row corresponds to a single node (such as a title or chapter) for a specific year and title. Table 3 provides a detailed description of each field included in this file. This metadata enables users to map and interpret the hierarchical structure of the U.S. Code.

| Field Name | Description | Example |
|---|---|---|
| Year | Edition year of the U.S. Code | 1926 |
| Title | Title number | 1 |
| Raw Text Length | Character count of the raw, unprocessed statutory text | 7,399 |
| Raw Text Word Count | Word count of the raw, unprocessed statutory text | 1,105 |
| Raw Text Unique Word Count | Number of unique words in the raw, unprocessed statutory text | 324 |
| Cleaned Text Length | Character count after cleaning (removing headers, OCR errors, etc.) | 6,724 |
| Cleaned Text Word Count | Word count after cleaning | 1,003 |
| Cleaned Text Unique Word Count | Unique word count after cleaning | 295 |
| Title Name | Human-readable name of the title | General Provisions |
| Entropies | Shannon entropy of the cleaned word frequency distribution | 7.16 |
| Entropies_norm | Normalized entropy (0–1 scale) | 0.87 |
| Scaling_Exponent_Heaps_law | Heaps' law exponent (vocabulary growth scaling) | 0.56 |
| Scaling_Exponent_ Zipf_Mandelbrot_law | Zipf-Mandelbrot law exponent (rank-frequency scaling) | 0.90 |
| Group | Functional grouping of the title (e.g., Government Structure, Civil Rights, etc.) | Government Structure |

**Table 1:** Fields and descriptions for Dataset 1: Contents of U.S. Code

| Field Name | Description | Example |
|---|---|---|
| Year | Edition year of the U.S. Code | 1994 |
| Title | Title number | 1 |
| Title Name | Human-readable name of the title | General Provisions |
| Average_Depth | Mean depth of nodes in the tree | 2.32 |
| Max_Depth | Maximum depth (longest path from root to leaf) | 4 |
| Average_Breadth | Mean number of nodes per hierarchical level | 12.6 |
| Max_Breadth | Maximum number of nodes at any single hierarchical level | 37 |
| Num_Leaf_Nodes | Number of leaf nodes (nodes with no children) | 53 |
| Average_Branching_Factor | Mean number of children per non-leaf node | 6.2 |
| Variance_Branching_Factor | Variability in the number of children across non-leaf nodes | 23.96 |
| Number_of_Nodes | Total number of nodes in the tree (all hierarchical units) | 63 |
| Group | Functional grouping of the title (e.g., Government Structure, Civil Rights, etc.) | Government Structure |

**Table 2:** Fields and descriptions for Dataset 2: Summary Statistics of hierarchical Structure

## Dataset 3: Cross-Reference Relationship

Data Set 3 captures the network of explicit cross-references (citations) between statutory provisions in the U.S. Code, allowing for detailed analysis of how sections, chapters, and titles are interconnected over time. Citation data up to 1988 were extracted from OCR-processed statutory text, while data from 1994 onward were obtained through web crawling of official government sources. The `Cross_Reference.csv` file covers the entire period from 1926 to 2023. Additionally, `edge_list_ocr.csv` contains OCR-extracted data from 1926 to 2000, and `edge_list_web.csv` contains web-crawled data from 1994 to 2023. All three CSV files follow the same structure, as described in Table 4, providing each citing–cited pair along with the citation context and relevant metadata, though the `Cross_Reference.csv` file excludes the `Citing_text` column.

| Field Name | Description | Example |
|---|---|---|
| Node Label | Unique identifier for the node within the hierarchy (e.g., Title, Chapter, Section) | chapter 1 |
| Corresponding Name | Human-readable name or description of the node | RULES OF CONSTRUCTION |
| Title | U.S. Code title number to which the node belongs | 1 |
| Year | Edition year of the U.S. Code in which this node appears | 1994 |

**Table 3:** Fields and descriptions for Dataset 2: Node Metadata for Hierarchical Structure

| Field Name | Description | Example |
|---|---|---|
| Citing | The title number of the citing provision (the source of the citation) | 2 |
| Cited | The title number of the cited provision (the target of the citation) | 5 |
| Year | The year of the U.S. Code edition in which the citation appears | 1926 |
| Weight | The number of times this specific citation occurs in the given year (edge weight) | 2 |
| Citing_text | The cleaned sentence from the citing provision containing the reference to the cited title, providing the context of the citation | ".... be paid to such employees be construed as a double salar y under the provisions of section 93 of title 5." |
| Citing Title | The human-readable name of the citing title | The Congress |
| Cited Title | The human-readable name of the cited title | Government Organization and Employees |
| Citing Group | The functional group or category of the citing title | Government Structure |
| Cited Group | The functional group or category of the cited title | Government Structure |

**Table 4:** Fields and descriptions for Dataset 3: Node Metadata for Cross-Reference Relationship

To highlight the most significant citation relationships, we extract a statistical backbone from the full cross-reference network using the disparity filter method [26]. The resulting backbone network is provided in `Backbone_structure.csv`, which follows the same format as described in Table 4.

# Technical Validation

To assess the consistency and reliability of our reconstructed *U.S. Code* dataset, we conduct technical validation along two dimensions: external validation via benchmarking data and internal validation via OCR–web comparison.

## External Validation via Benchmarking

We benchmark our dataset against the corpus used by Coupette et al. (2021) [14], which analyzes the evolution of statutory and regulatory texts in the United States using a network-based framework. Their research is based on the official XML or XTHML versions of the U.S. Code, available from government sources [33, 34]. While we draw from the same underlying sources, our dataset differs in format and scope. We use both digitized and OCR-reconstructed textual versions of the U.S. Code, spanning from 1934 to 2023, including editions that are not available in machine-readable format. The pre-1988 volumes were recovered using OCR combined with generative AI-based processing to ensure consistency and completeness across the full historical record.

To validate our reconstructed dataset, we compare it with the benchmark dataset across three dimensions: (1) word count per title (Fig. 11B), (2) the number of structural nodes in the hierarchical tree of each title (Fig. 11D), and (3) the number of inter-title reference links (in-degree and out-degree) of each titles (Fig. 11E). Across all three dimensions, we observe strong positive correlations, demonstrating that our dataset is highly consistent with the benchmark data.

Despite overall alignment in trends, several methodological differences account for discrepancies in the numerical values. Unlike the benchmark dataset, which includes only the statutory body text (i.e., black-highlighted sections in Fig. 11A), our dataset retains all explicit components of the U.S. Code—including headers, historical notes, and revision records—which constitute approximately 54% of the total word count per title (Fig. 11C). Furthermore, for the hierarchical structure of each title, we adopt a more conservative parsing strategy for under section level such as subsections and items, identifying only those that conform to rule-based drafting patterns. This results in fewer, but more structurally coherent, nodes in the hierarchical structure. Lastly, our reference extraction includes cross-title citations and references found in supplementary materials, contributing to higher reference counts. These differences reflect our aim to construct a more comprehensive and reusable dataset for future legal and computational research. Specifically, by retaining all textual components of the U.S. Code—not just statutory provisions but also headers, historical notes, and revision records—we enable future researchers to decide whether to focus solely on the statutory content or to include non-statutory materials when analyzing the broader evolution of the Code. This flexibility supports a wide range of research objectives, from narrow legal analysis to studies of institutional or linguistic change in the legislative corpus.

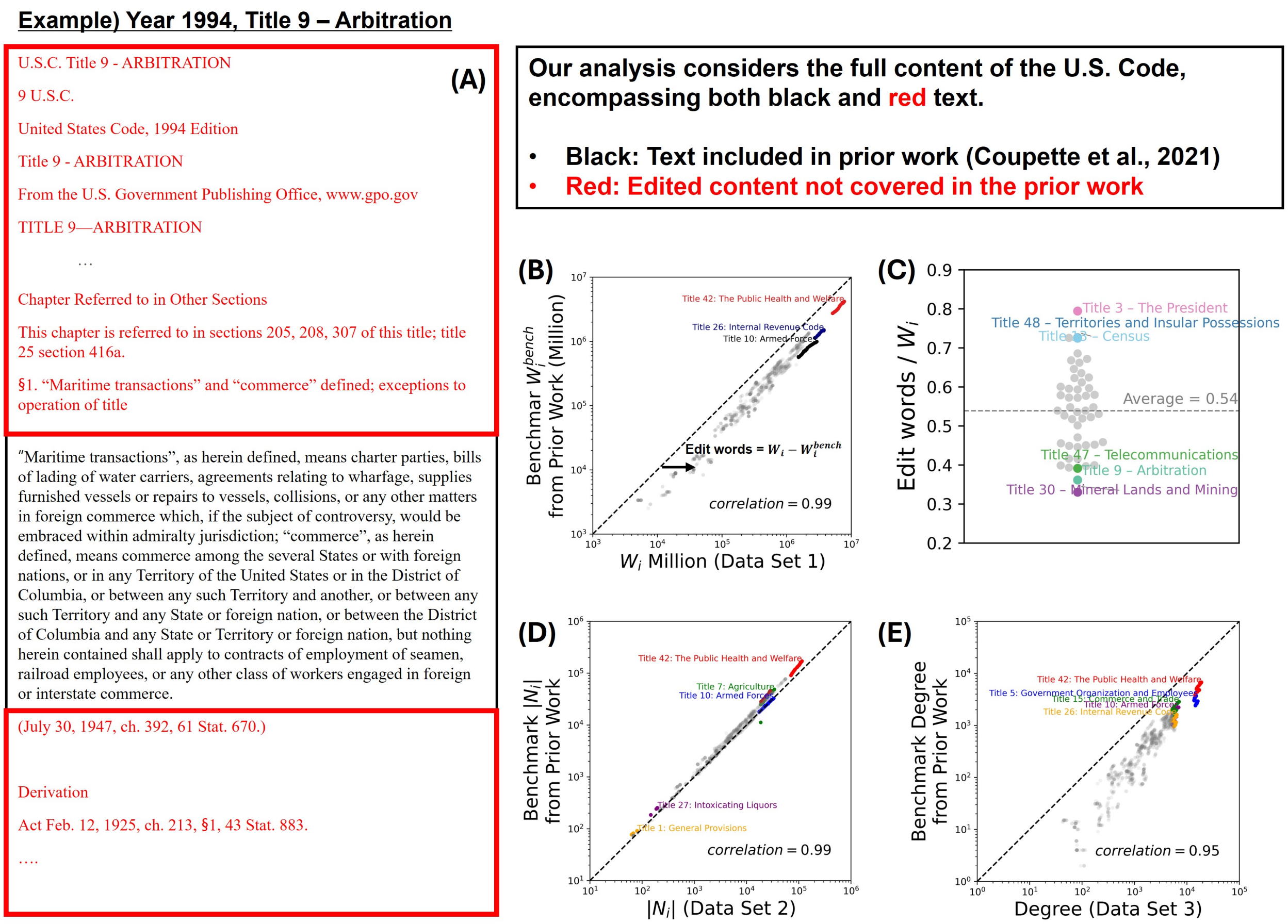

**Figure 11: Comparison with prior work and technical validation of our U.S. Code dataset.** (A) Example of the U.S. Code text in Title 9 (Arbitration) in 1994. The black text corresponds used in the prior dataset (Coupette et al., 2021), while our dataset includes the entire text of the U.S. Code, encompassing both black and red text. The red text represents broader contextual information such as amendments, revision note which are not covered in the prior work. (B) Comparison of total word counts between our dataset ($W_i$) and the prior benchmark ($W_i^{\text{bench}}$), showing a near-perfect correlation ($r = 0.99$). (C) Proportion of additional words in our dataset not included in the prior work, computed as $(W_i - W_i^{\text{bench}})/W_i$. On average, over 54% of the additional words in each title are included, highlighting the substantial amount of previously unaccounted information. (D) Comparison of the number of structural nodes ($|N_i|$) in the hierarchical tree of each title. While prior work counts nodes at the paragraph and subparagraph levels, our method uses more refined structural units (e.g., items, subitems), showing a high correlation ($r = 0.99$). (E) Network degree (in-degree + out-degree) comparison in the cross-reference network among titles. As our dataset includes additional references embedded in the red text, it captures a larger number of legal cross-links ($r = 0.95$).

## Internal Validation via OCR–Web Comparison

To further validate the accuracy of our OCR-based reconstruction of early editions of the U.S. Code (prior to 1994), we compare the OCR-reconstructed versions of the 1994 and 2000 edition with their corresponding official web-based versions. We assess consistency across four dimension: (1) total word count for Title ($W_i$), (2) unique word count for Title ($V_i$), (3) edge weight of citation links between Titles, and (4) distribution of difference in edge weight.

Across all three quantitative measures—word count, unique word count, and edge weight—we observe very high agreement between the OCR and web datasets, with data points closely aligning along the $y = x$ line and $R^2$ values approaching 1. This indicates that the reconstructed documents match the web versions with a high degree of fidelity. Additionally, the distribution of citation weight differences ($\Delta$ edge weight) is tightly centered around zero, suggesting near-perfect agreement in the structure of the citation network.

This internal validation confirms that our OCR pipeline, enhanced by generative AI processing, reliably reproduces both the structure and content of the official digital versions of the U.S. Code. Together with the external validation, this test demonstrates that our dataset satisfies both external consistency with benchmark corpora and internal reliability in digitizing historical legal texts.

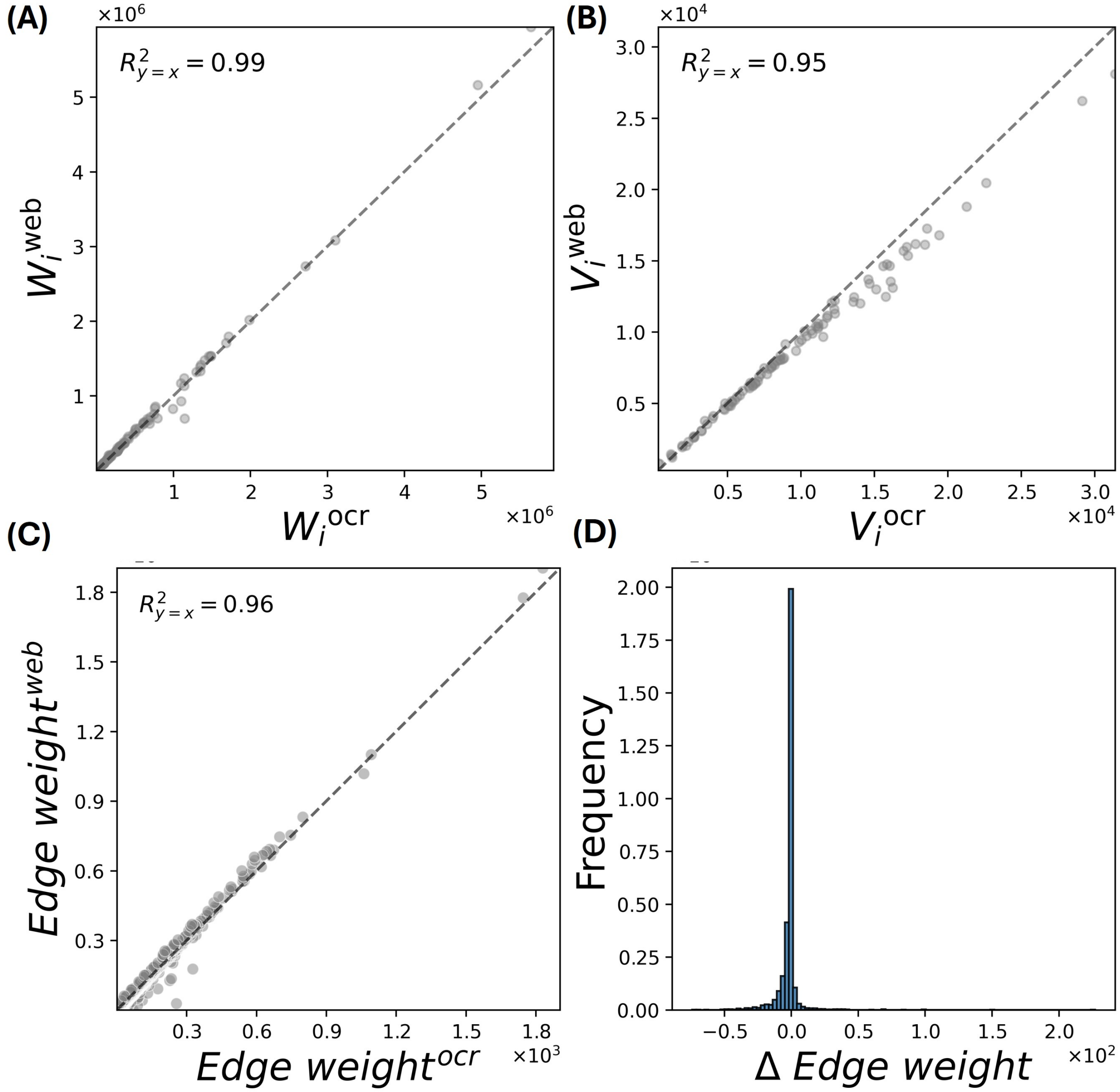

**Figure 12: Internal validation of the OCR-reconstructed U.S. Code dataset.** (A) Comparison of total word counts of each title between the OCR-reconstructed dataset ($W_i^{ocr}$) and the web version ($W_i^{web}$) for the years 1994 and 2000. The data show a near-perfect R-square with y=x line ($R^2 = 0.99$). (B) Comparison of unique word counts ($V_i$) across titles, also demonstrating high consistency ($R^2 = 0.95$). (C) Comparison of citation edge weights between titles, with strong alignment across sources ($R^2 = 0.96$). (D) Histogram of edge weight differences between the web and OCR datasets (Δ edge weight = web − OCR), indicating that most citations are perfectly matched (difference = 0).

# Usage Notes

The dataset provides a century-spanning, structured digital record of the United States Code (U.S. Code) from 1934 to 2023, enabling large-scale computational analysis of statutory evolution, legal language, hierarchical structure, and cross-references among legal domains. The following notes are intended to guide researchers in accessing, processing, and reusing the data effectively:

**Data Access and Formats:** The dataset is organized into three main components: (1) the summarized word content information of each title, (2) hierarchical tree structures representing the internal organization of legal provisions, and (3) cross-reference networks among titles. Data files are provided in widely used formats, including CSV (for word content, node-level and cross-reference metadata), and Python networkx objects (for hierarchical trees and network structures). Researchers can access these files directly from the repository specified in the Data Records section.

**Preprocessing and Data Quality:** For editions prior to 1994, the text was reconstructed from OCR scans and cleaned using generative AI models, with manual review to correct errors. Despite rigorous validation, users should be aware that OCR-derived text may contain residual artifacts, particularly in the earliest volumes (e.g., the 1926 edition, which is included for completeness but not analyzed due to unresolved errors). For years or titles with missing official digital records, data were interpolated using the nearest available edition to ensure temporal continuity. These interpolations are documented in the Methods section and should be considered when conducting longitudinal analyses.

**Hierarchical Structure Limitations:** Hierarchical tree data (Data Set 2) are available only from 1994 onward, as reliable extraction of structural information from pre-1994 OCR text was not feasible due to the loss of indentation and formatting cues during digitization. For analyses requiring full hierarchical depth, users should restrict their scope to the post-1994 period.

**Cross-Reference Networks:** The cross-reference data (Data Set 3) are constructed using regular expression-based pattern matching for explicit title references within the text. While this approach captures a broad range of legal interdependencies, it may not identify implicit or non-standard references. The backbone networks, filtered for statistical significance, are provided to facilitate network analysis at varying levels of granularity. Users interested in alternative backbone thresholds can apply the provided algorithms to the raw edge lists.

# Code Availability

The code used for generative AI processing, data generation, technical validation, and figure creation is freely available at this GitHub link.

# Acknowledgment

The authors would like to acknowledge the support of the National Science Foundation Grant Award Number 2133863. H.Y. and J.Y. acknowledge Global Humanities and Social Sciences Convergence Research Program through the National Research Foundation of Korea (NRF), funded by the Ministry of Education (2024S1A5C3A02042671). H.Y. acknowledges the support from the Institute of Management Research at Seoul National University and the Emergent Political Economies grant from the Omidyar Network through the Santa Fe Institute. J.H. acknowledges the support of a Lou Schuyler grant from the Santa Fe Institute.

# Author Contributions

J.H. conceived the study, with input from H.Y. and D.J. J.H. and H.Y. supervised and coordinated the project. H.Y., C.K., and G.B.W. acquired funding for the study. J.H. performed initial collection and analyses. D.J. conducted extensive analyses that include coding machine-learning to curate the text from OCR, and parsing the text to construct the hierarchical structure, and cross-referencing. D.J. produced all figures and performed the technical validation of the data. J.H., H.Y., and D.J. contributed to the writing of the manuscript. D.J. and J.Y. wrote the data records and usage notes. All authors reviewed and edited the manuscript, approved the final version for publication, and agreed to be held accountable for the work presented herein.

# Supplementary Information

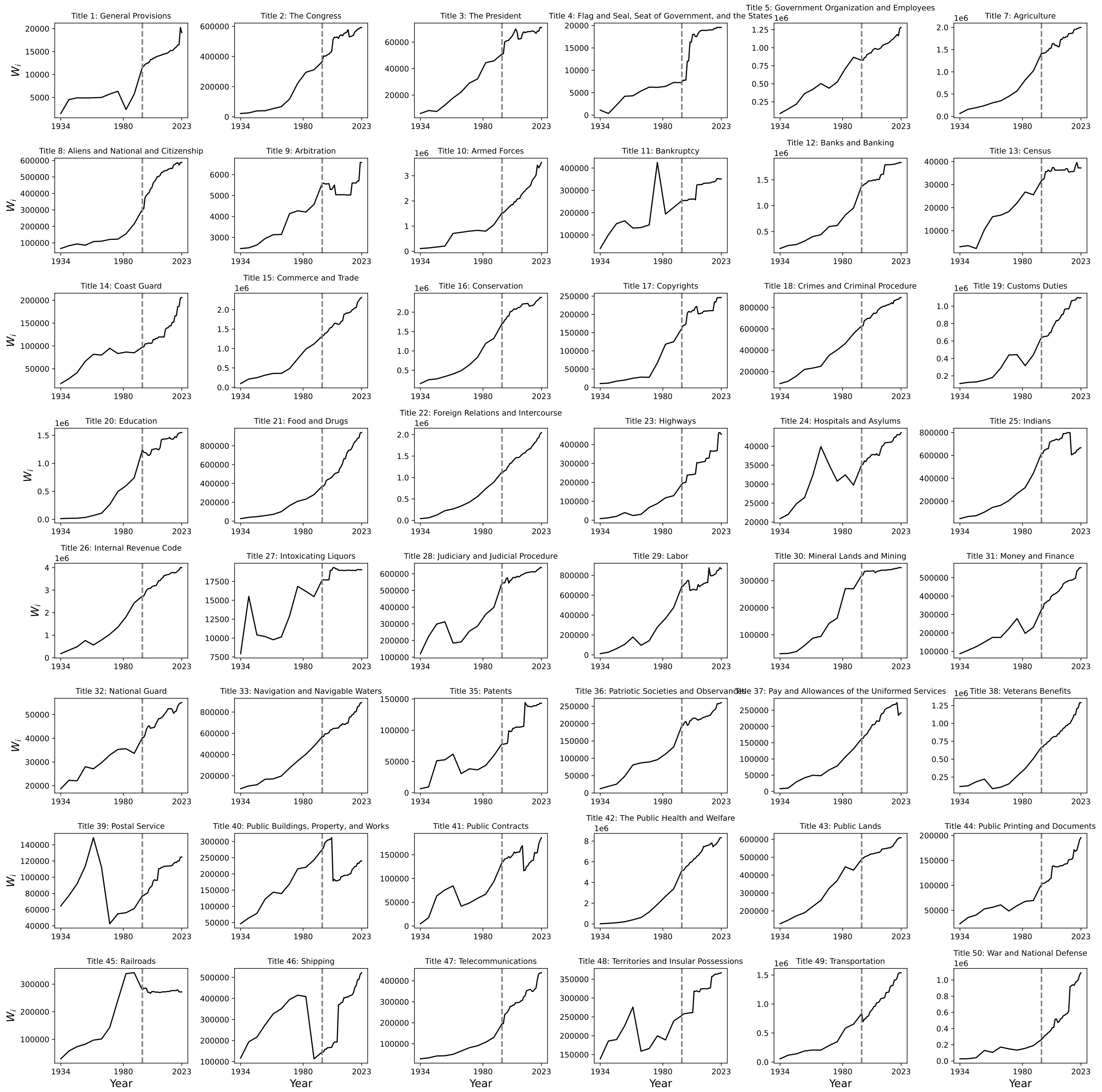

**Figure S1.1:** Word counts versus time across all titles.

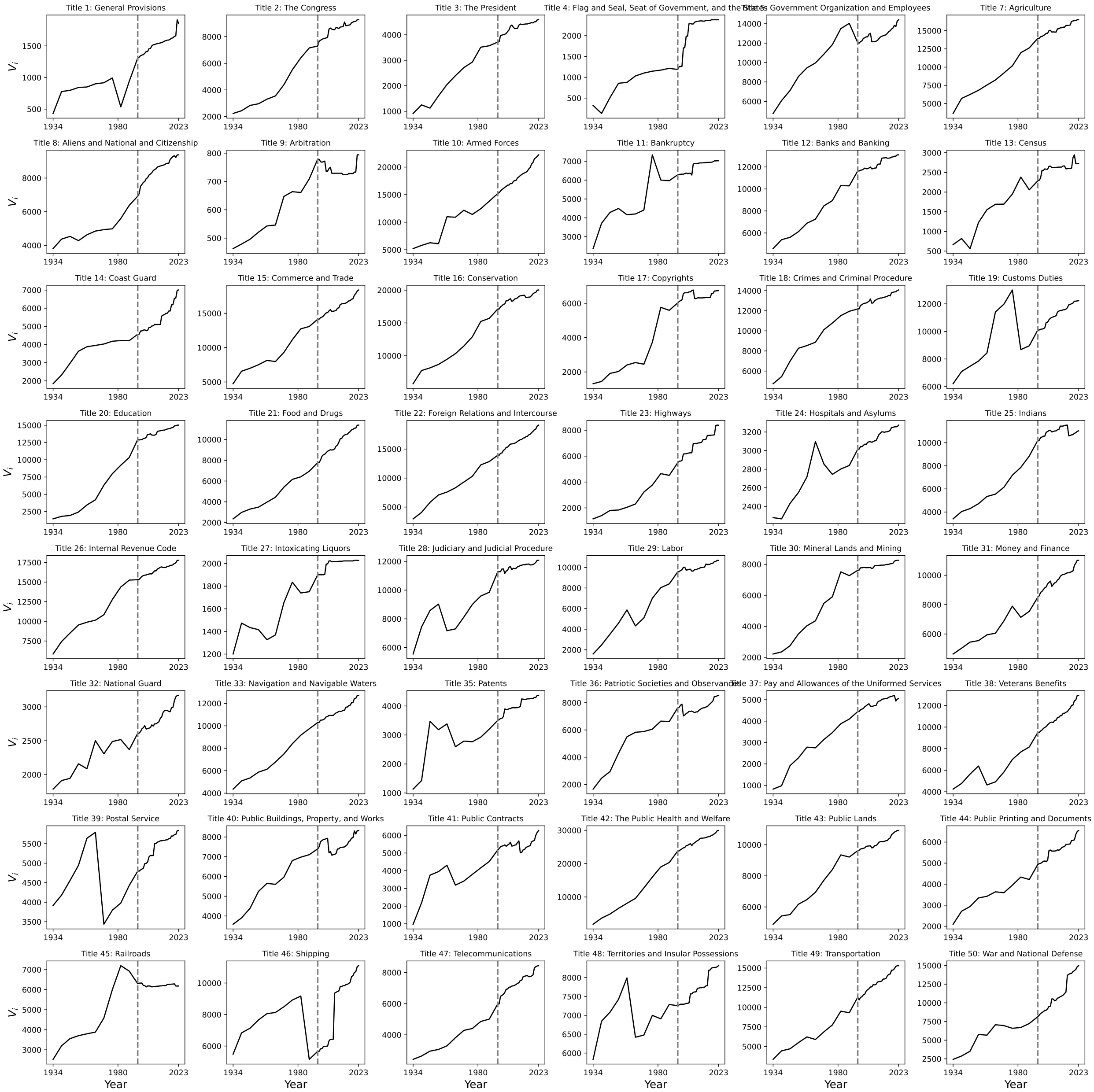

**Figure S1.2:** Unique word counts versus time across all titles.

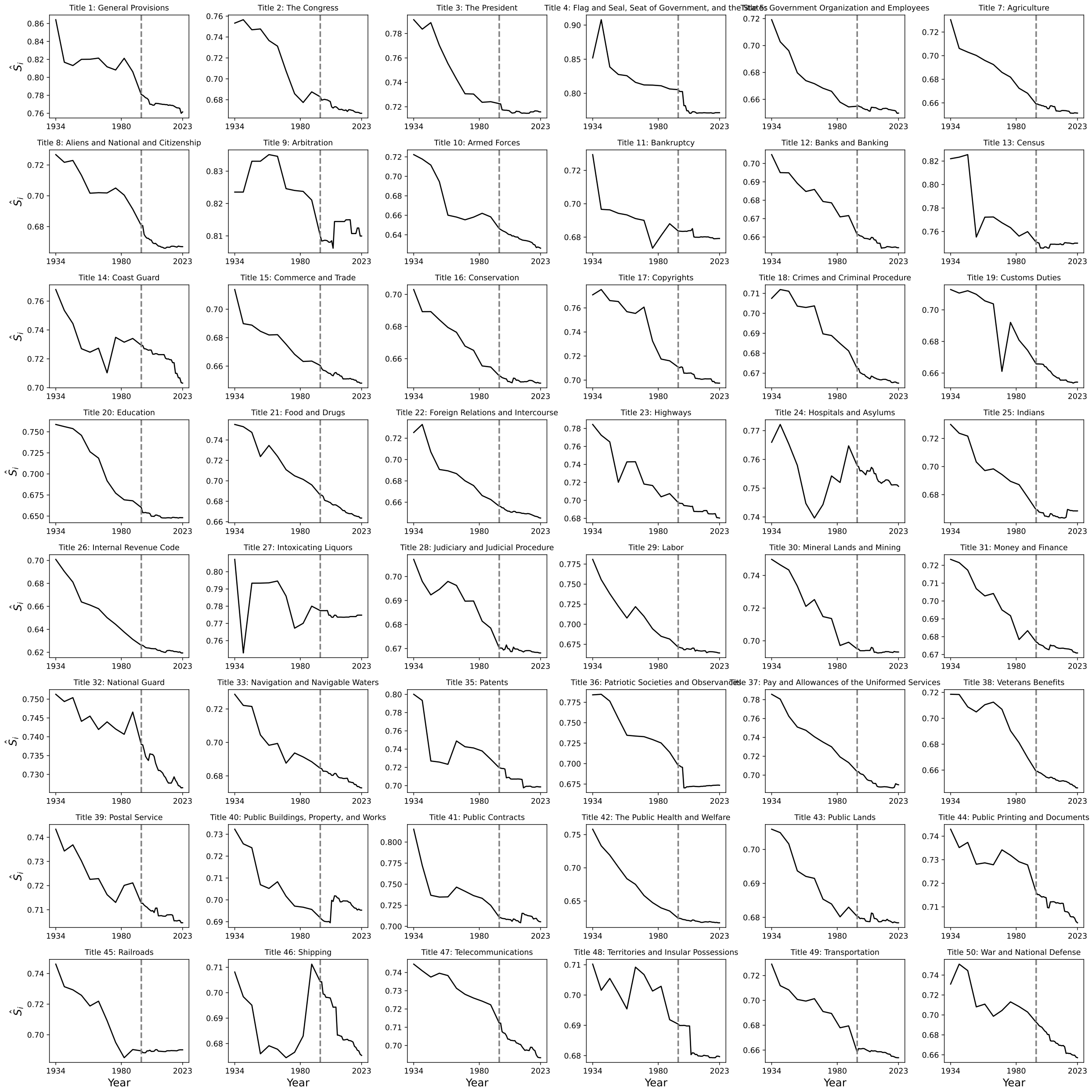

**Figure S1.3:** Normalized Shannon entropy across all titles.

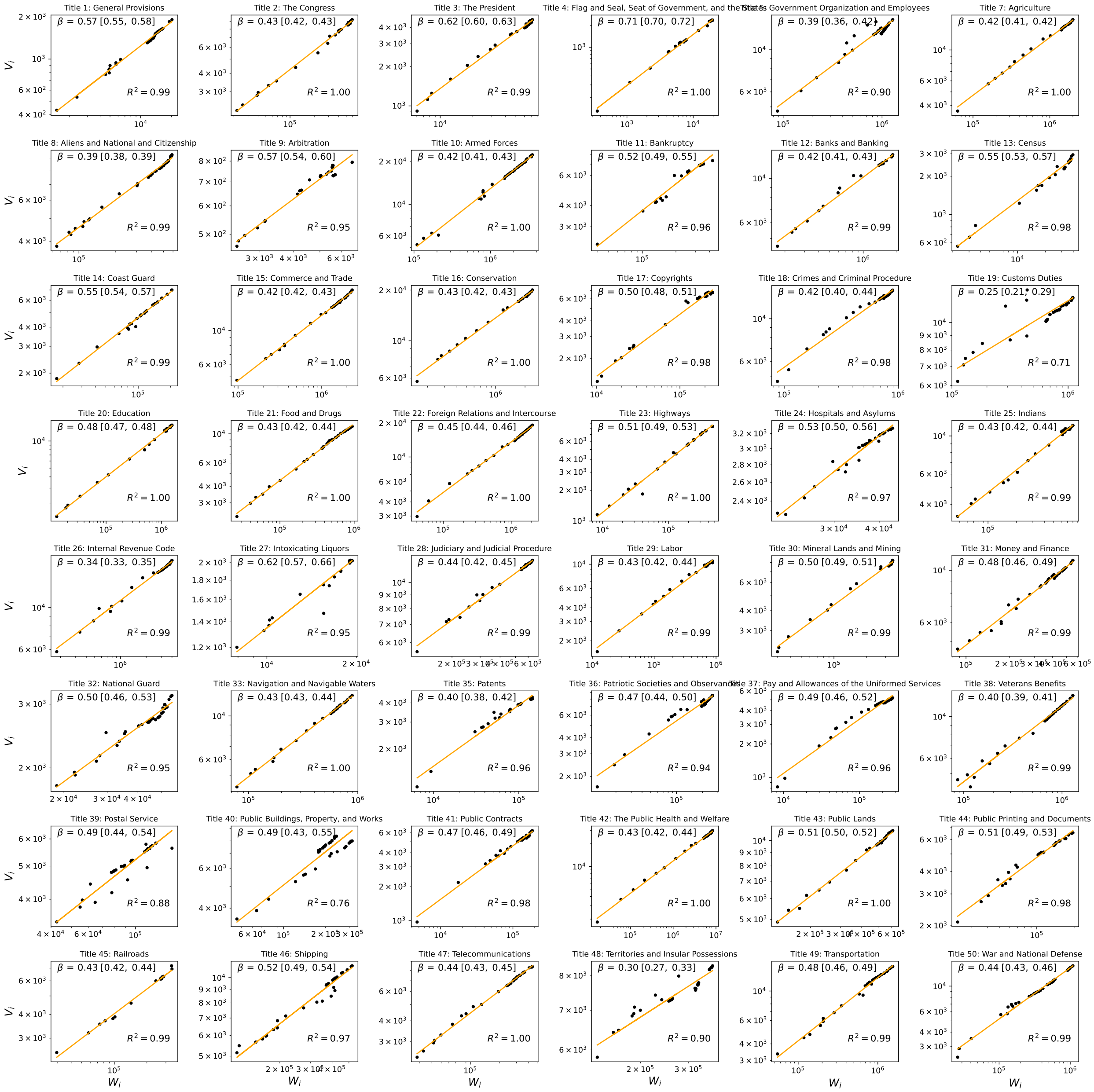

**Figure S1.4:** Heaps' law across all titles.

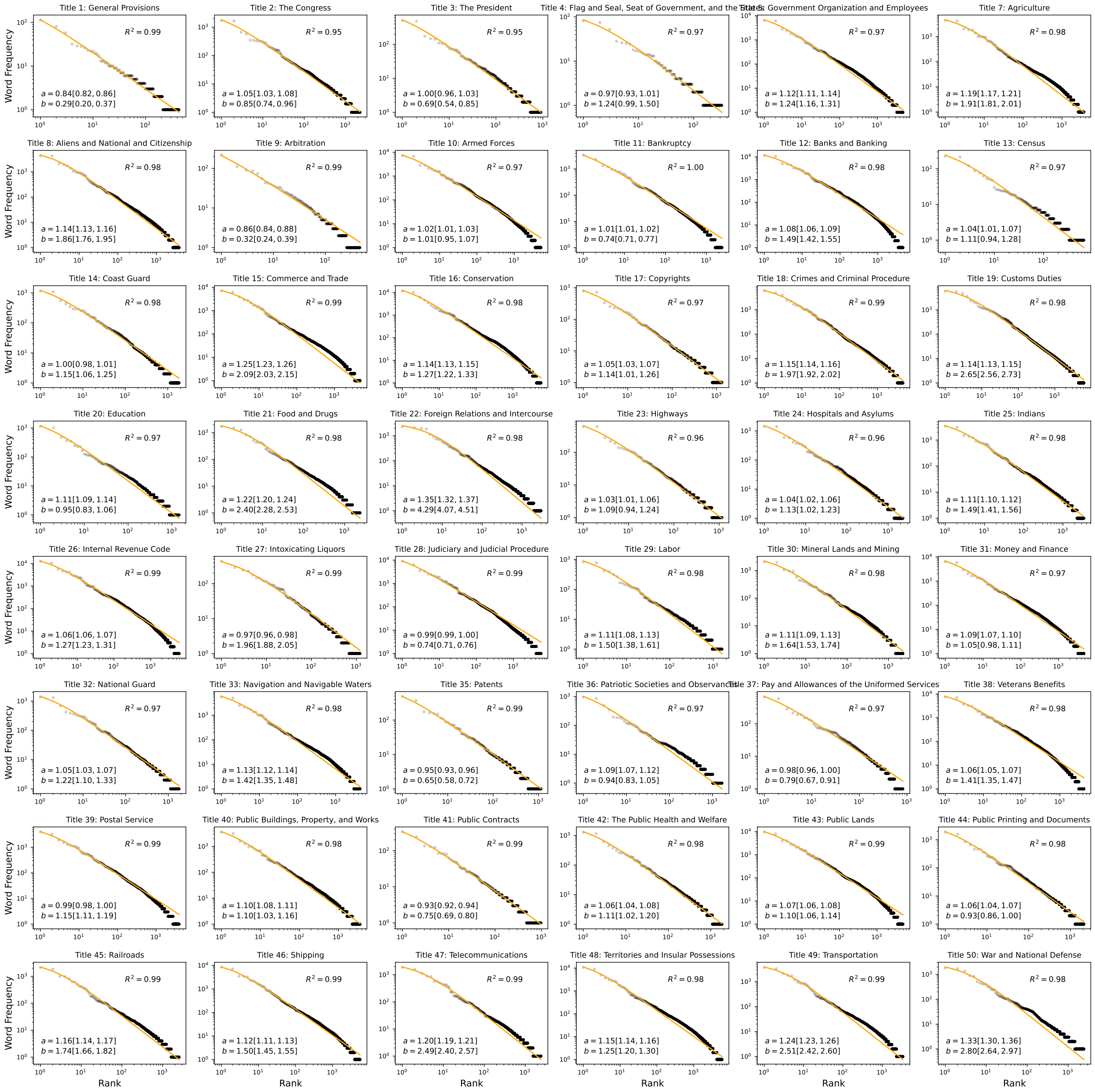

**Figure S1.5:** Zipf-Mandelbrot’ law across all titles in 1934.

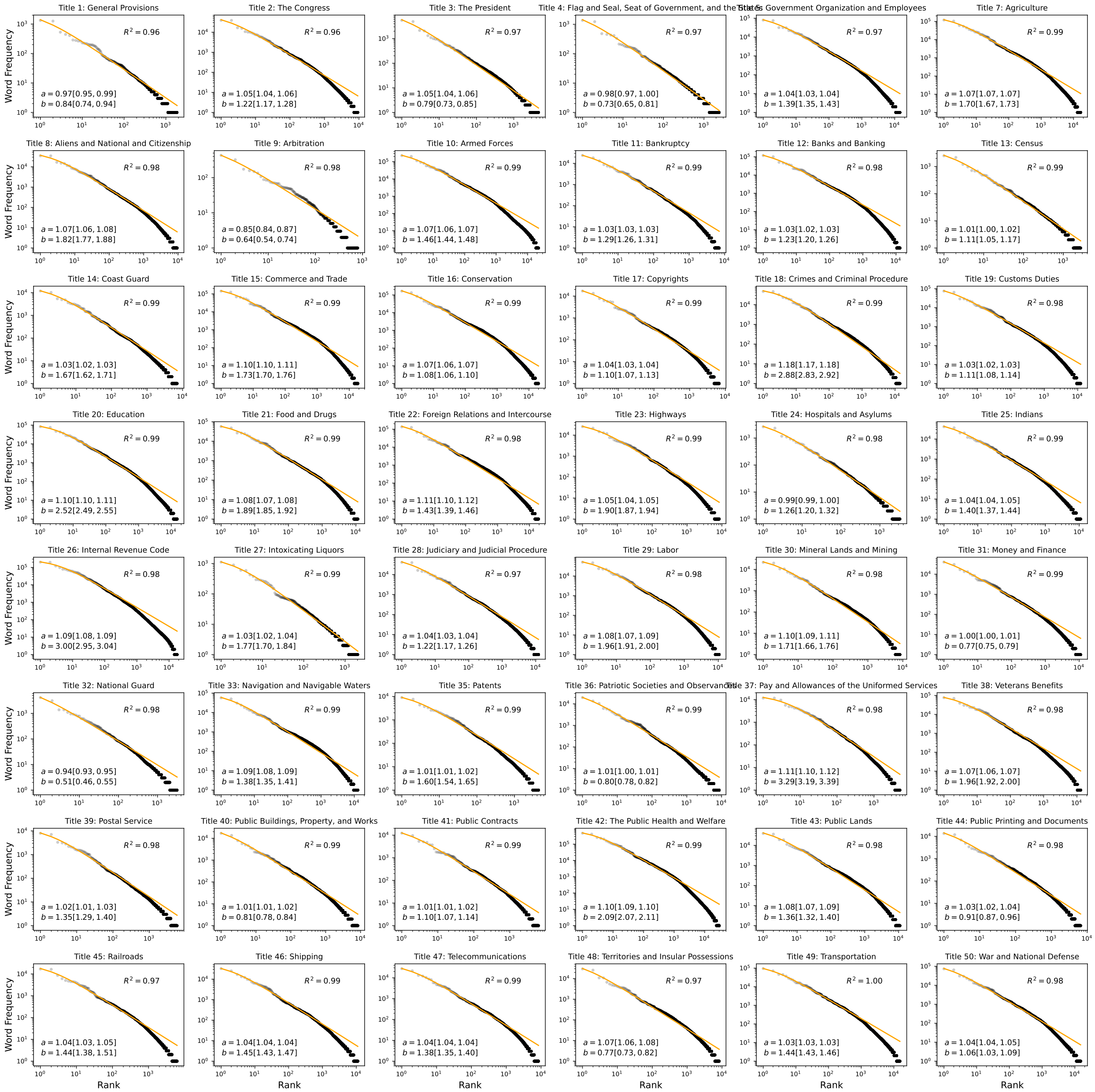

**Figure S1.6:** Zipf-Mandelbrot' law across all titles in 2023.

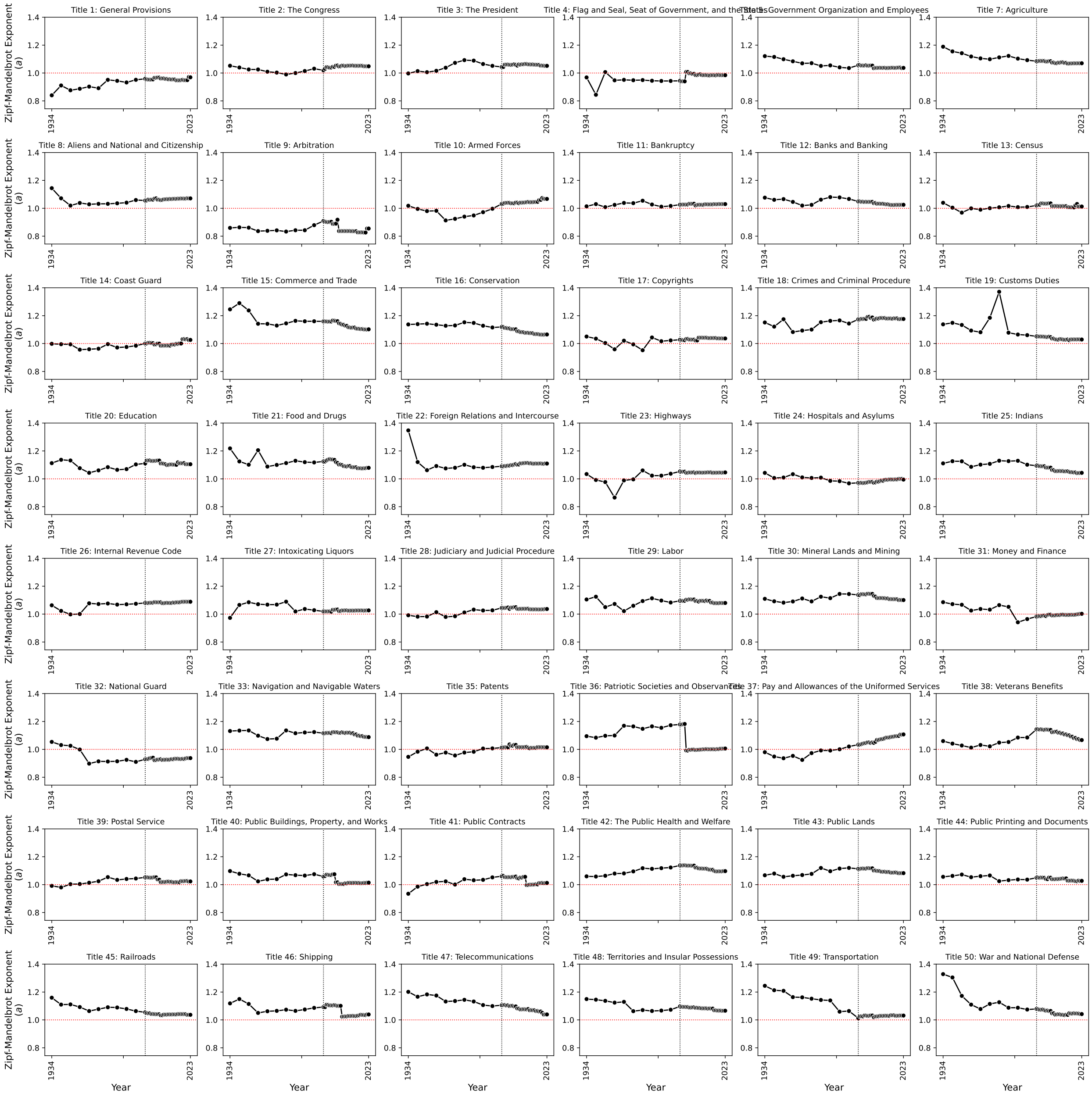

**Figure S1.7:** Zipf-Mandelbrot' law across all titles.

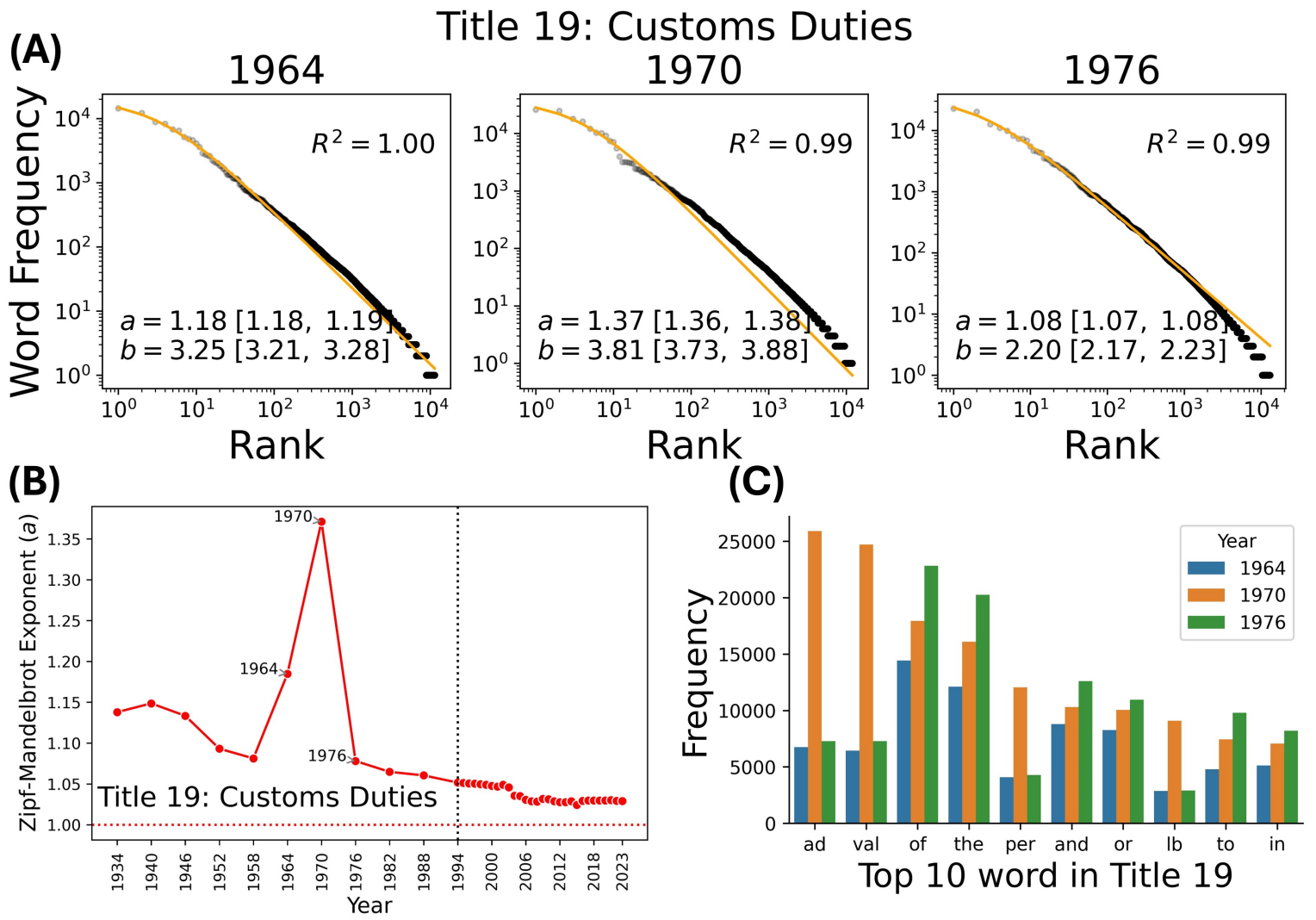

**Figure S1.8: Anomaly case in Zipf-Mandelbrot's law in Title 19.** The contents in Title 19 follow the Zipf-Mandelbrot law, exhibiting a power-law relationship between word frequency rank ($x$-axis) and word frequency ($y$-axis). (A) Zipf-Mandelbrot fits for Title 19 in 1964, 1970, and 1976; each scatter point represents a word in the corresponding year. (B) Yearly trend of the Zipf-Mandelbrot scaling exponent ($a$) for Title 19, highlighting a spike in 1970. (C) Top 10 most frequent words in Title 19 in 1970, compared with their corresponding frequencies in 1964 and 1976.

Title 19 shows a noticeable spike in the scaling exponent Zipf-Mandelbrot scaling exponent in 1970 (Fig. S1.8(B)). This deviation is not incidental: 1970 corresponds to a pivotal period in U.S. customs law, when the Tariff Schedules of the United States (TSUS) were newly integrated as a formal legal framework following the Kennedy Round of GATT negotiations. During this time, it was legally necessary to specify the validity and transitional schedule of tariff rates directly in Title 19. As a result, certain printed versions of Title 19 included extensive tables detailing these schedules, leading to a temporary surge in tabular data within the legal text (see Fig. S1.8(C)). This case illustrates how abrupt lexical shifts in the data often correspond to substantive legal and historical developments, affirming the fidelity of our OCR-processed corpus. Moreover, although some local anomalies like this exist, we observe that the overall average scaling exponent ($a$) exhibits a gradual decreasing trend over time. Notably, the data show no discontinuity around 1994—the point marking the transition between OCR-processed and web-based text sources—demonstrating the continuity and reliability of the dataset across different pre-processing regimes.

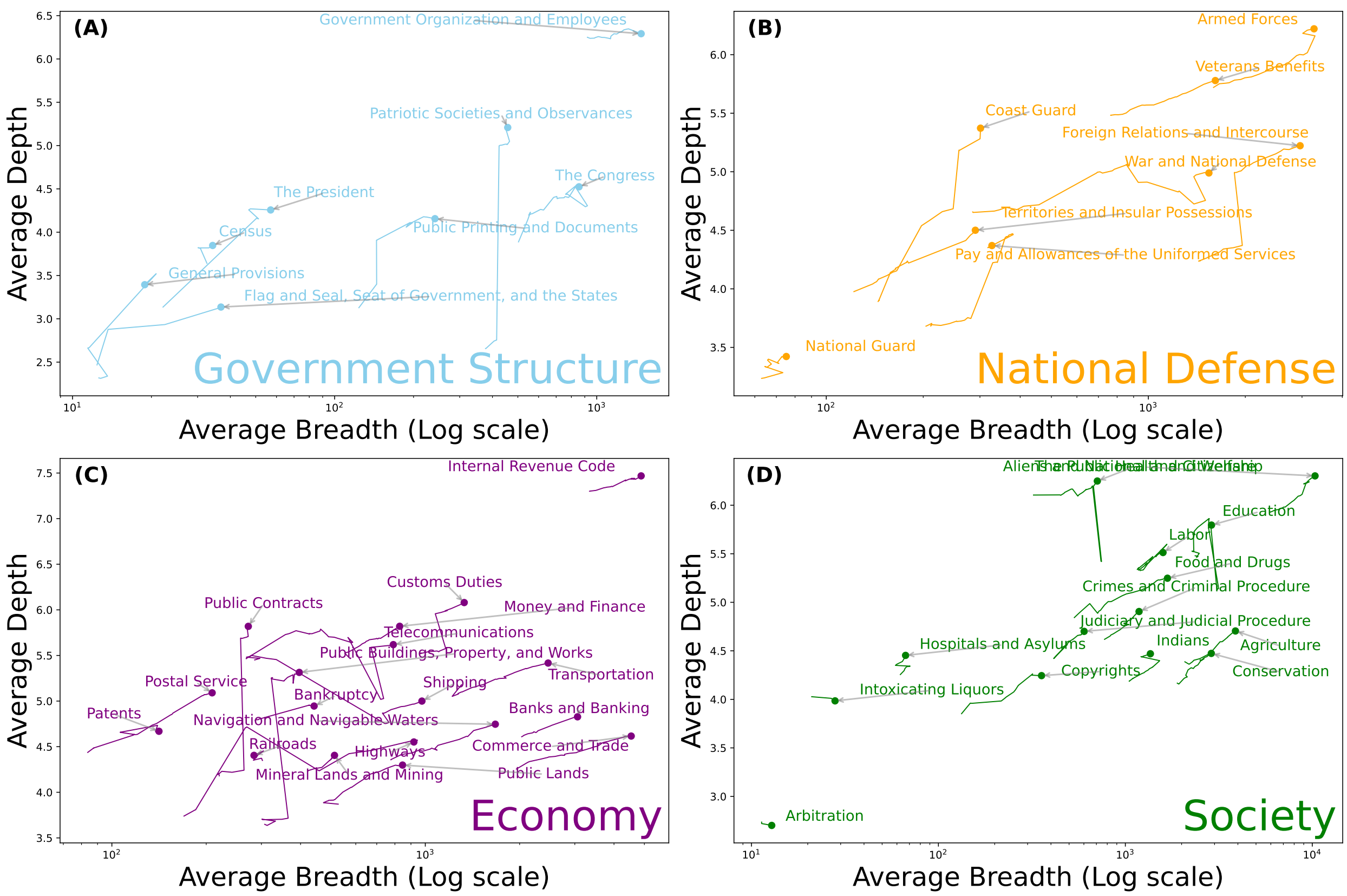

**Figure S1.9:** Structural Evolution of U.S. Code (linear-log relationship).

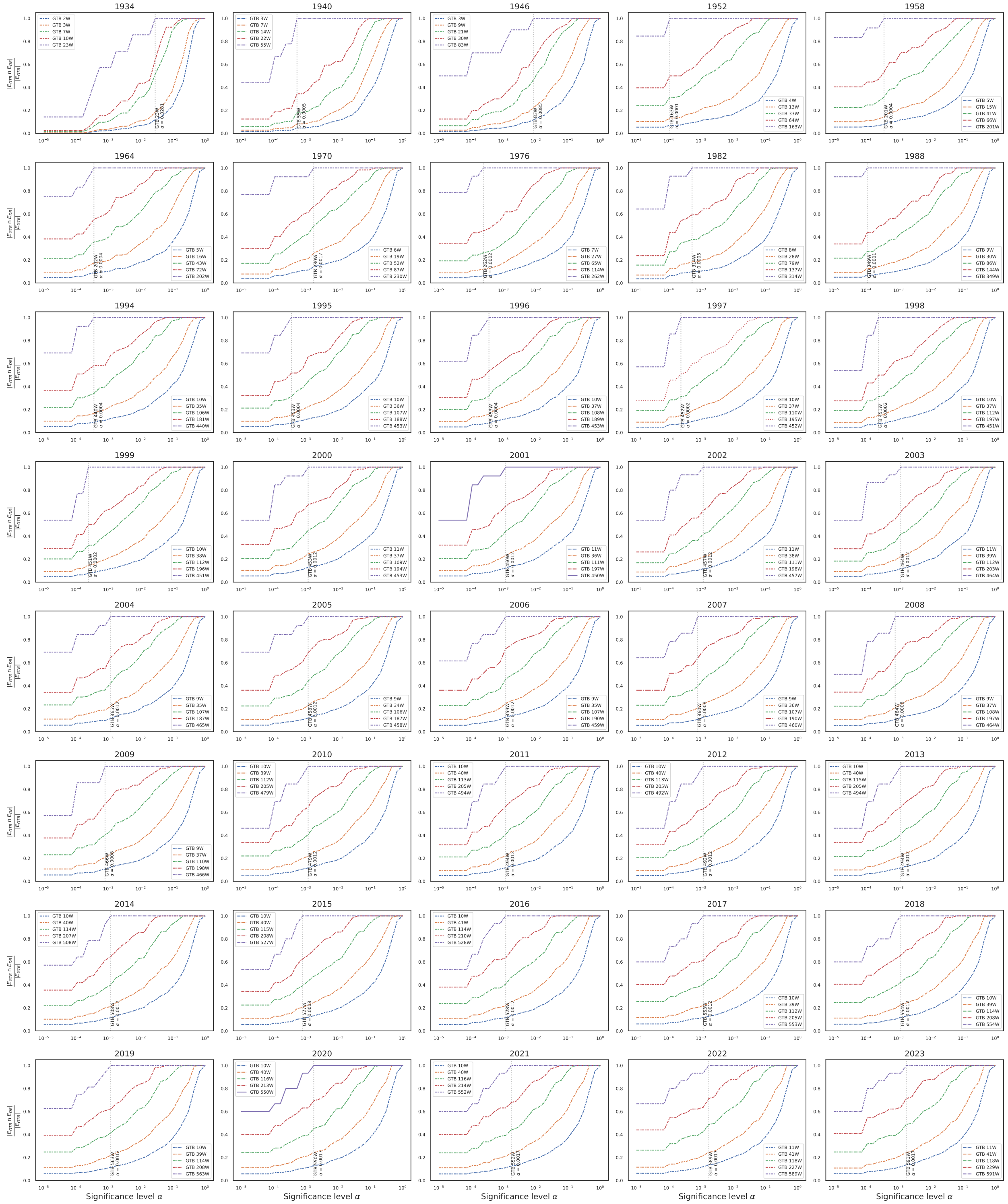

**Figure S1.10:** Alpha selection for backbone network.

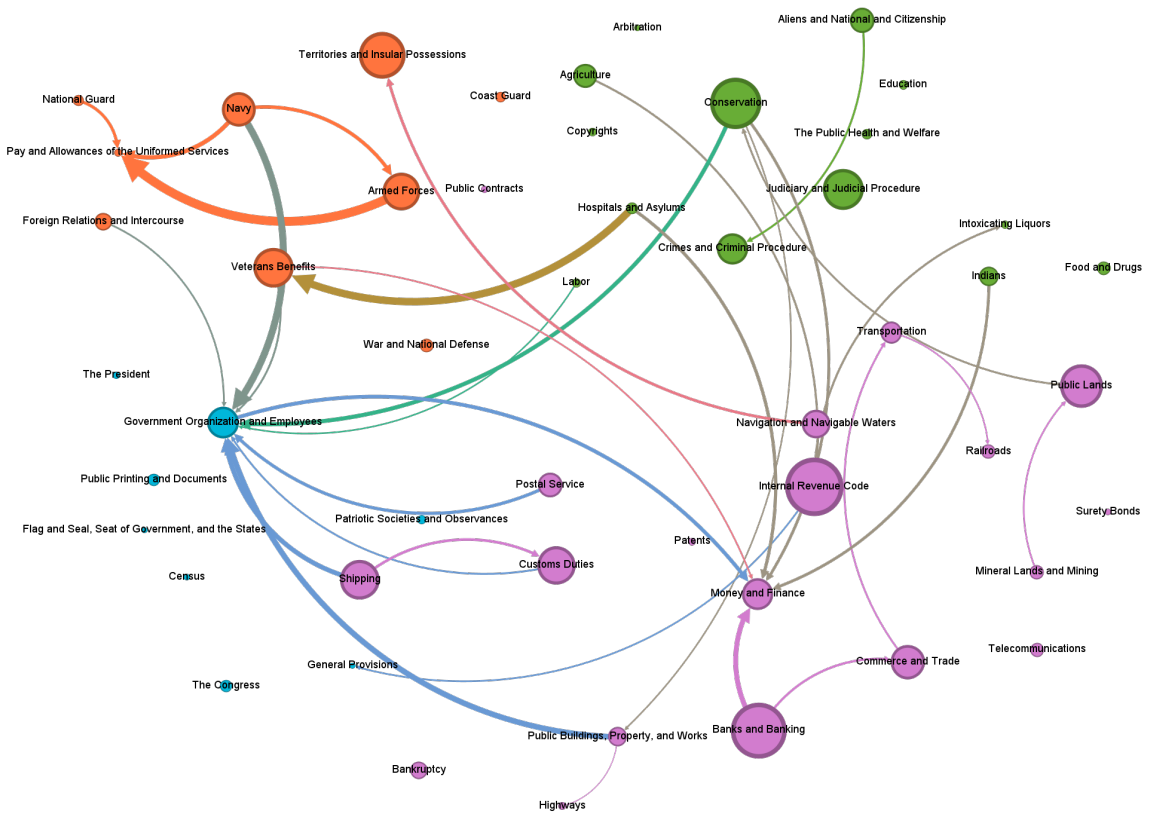

**(a)** year 1934

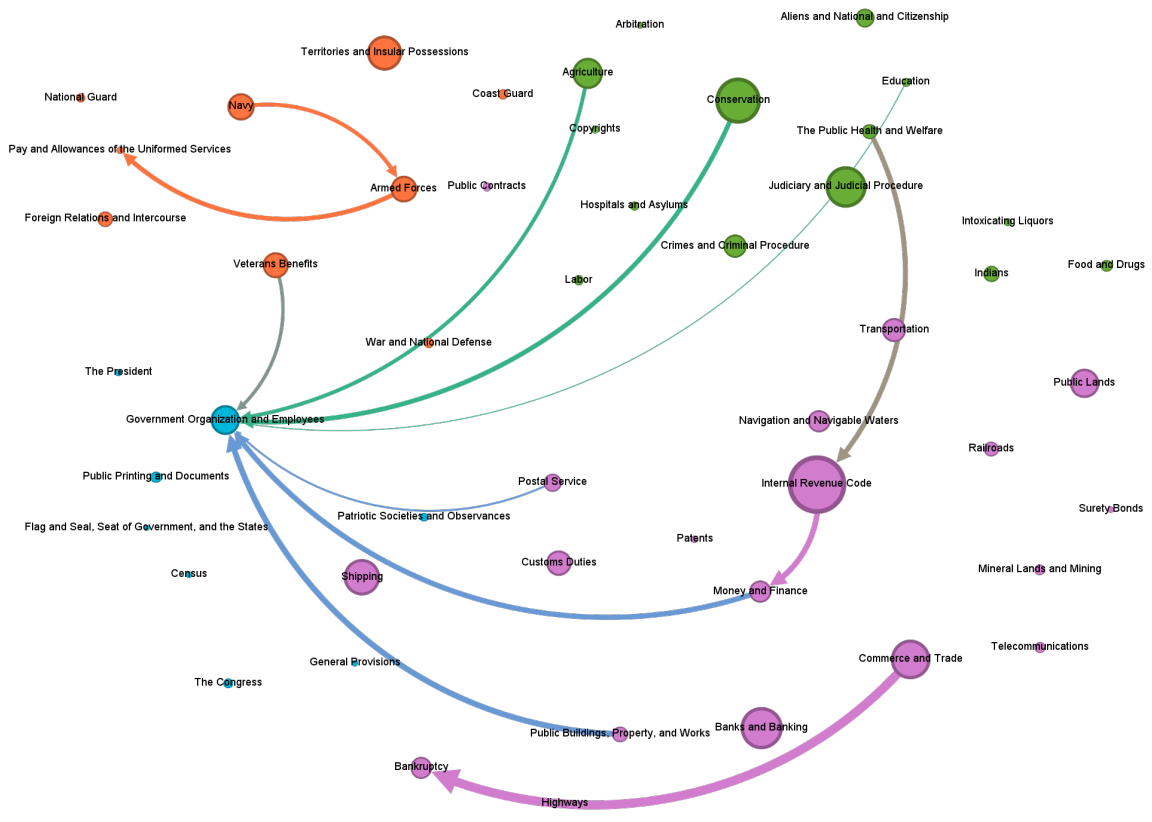

**(b)** year 1940

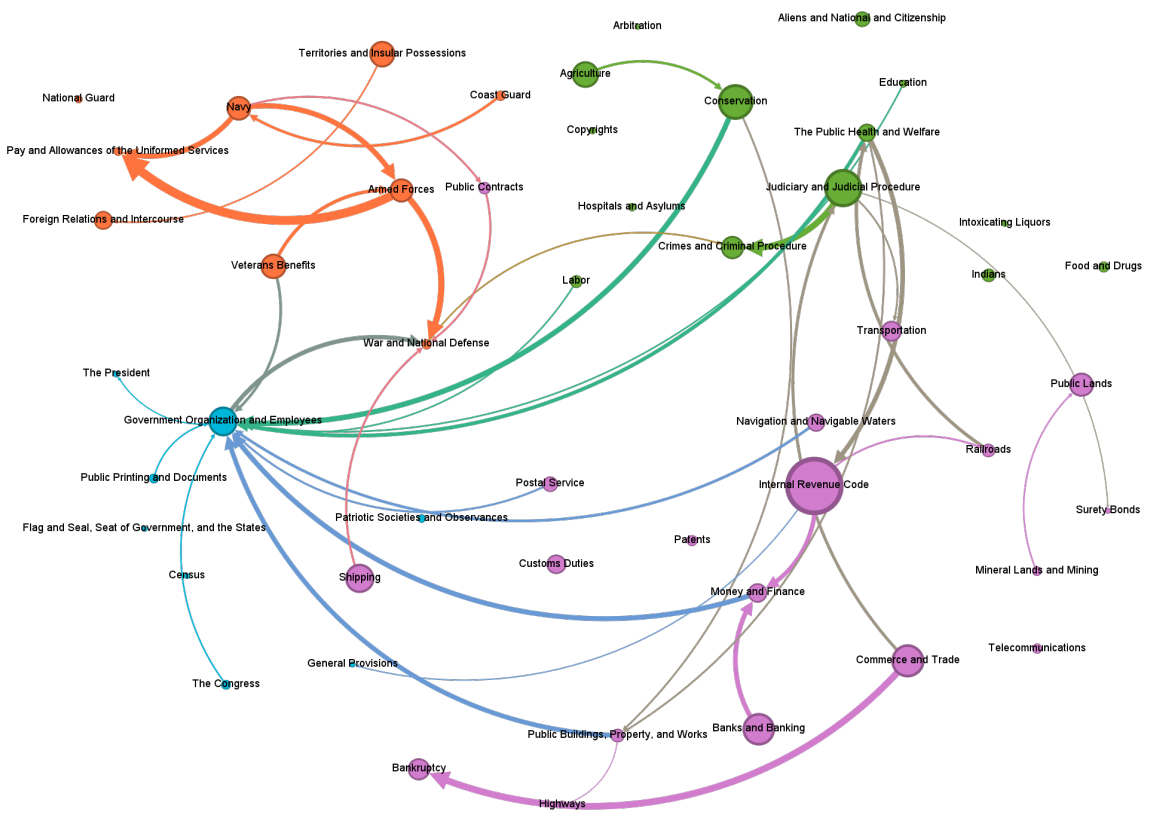

**(c)** year 1946

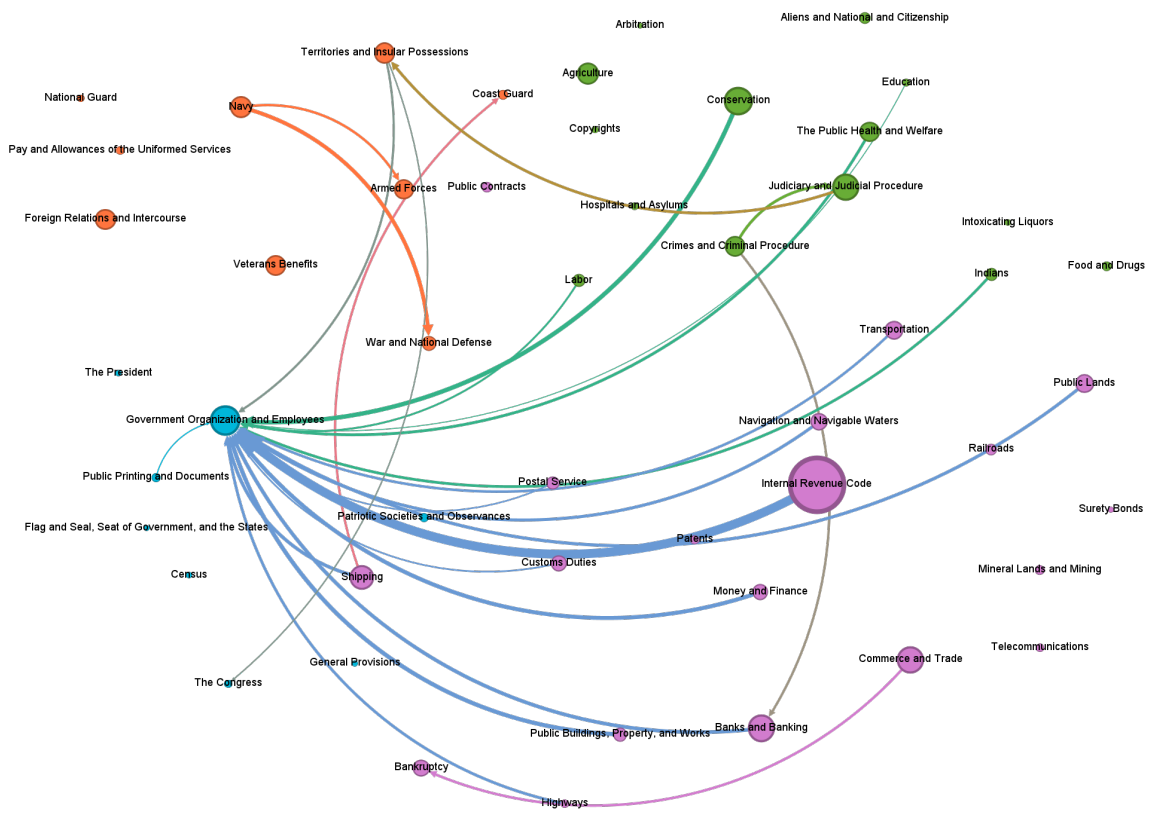

**(d)** year 1952

**Figure S1.11: Backbone structure of the U.S. Code's title cross-reference network in 1934, 1940, 1946, 1952.**

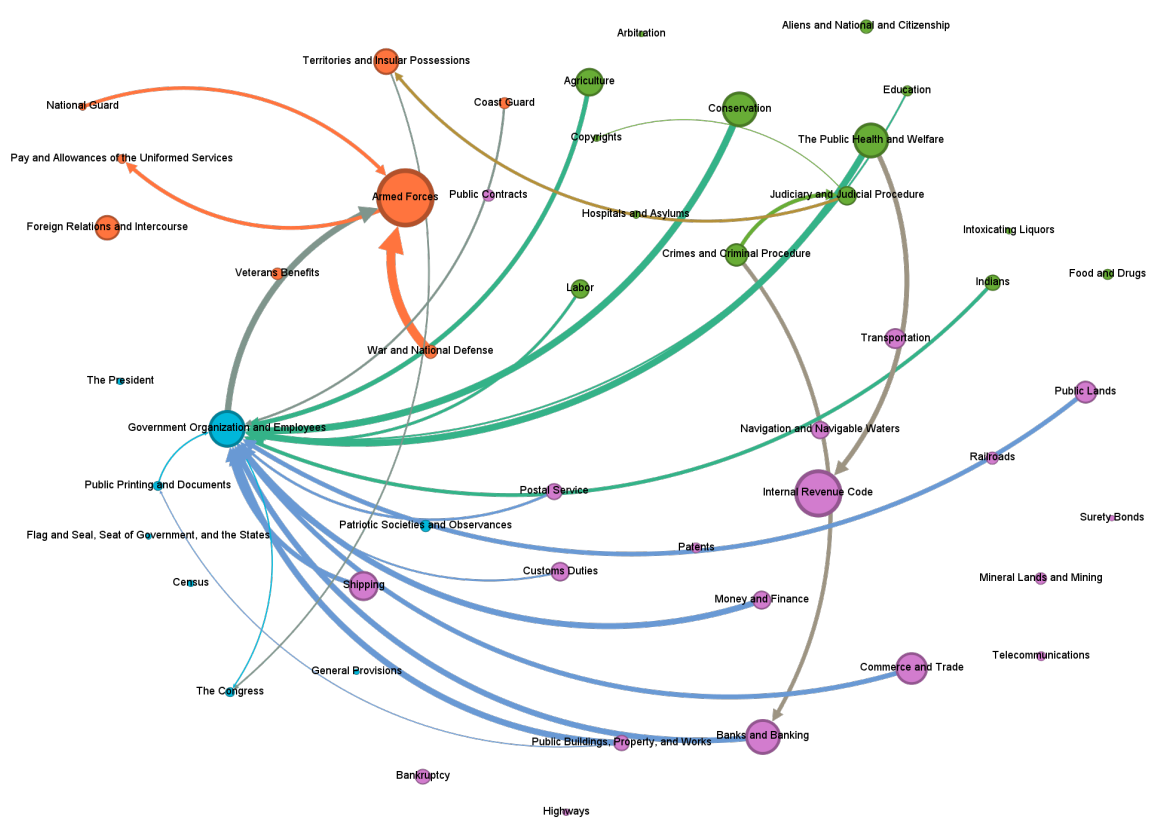

**(a)** year 1958

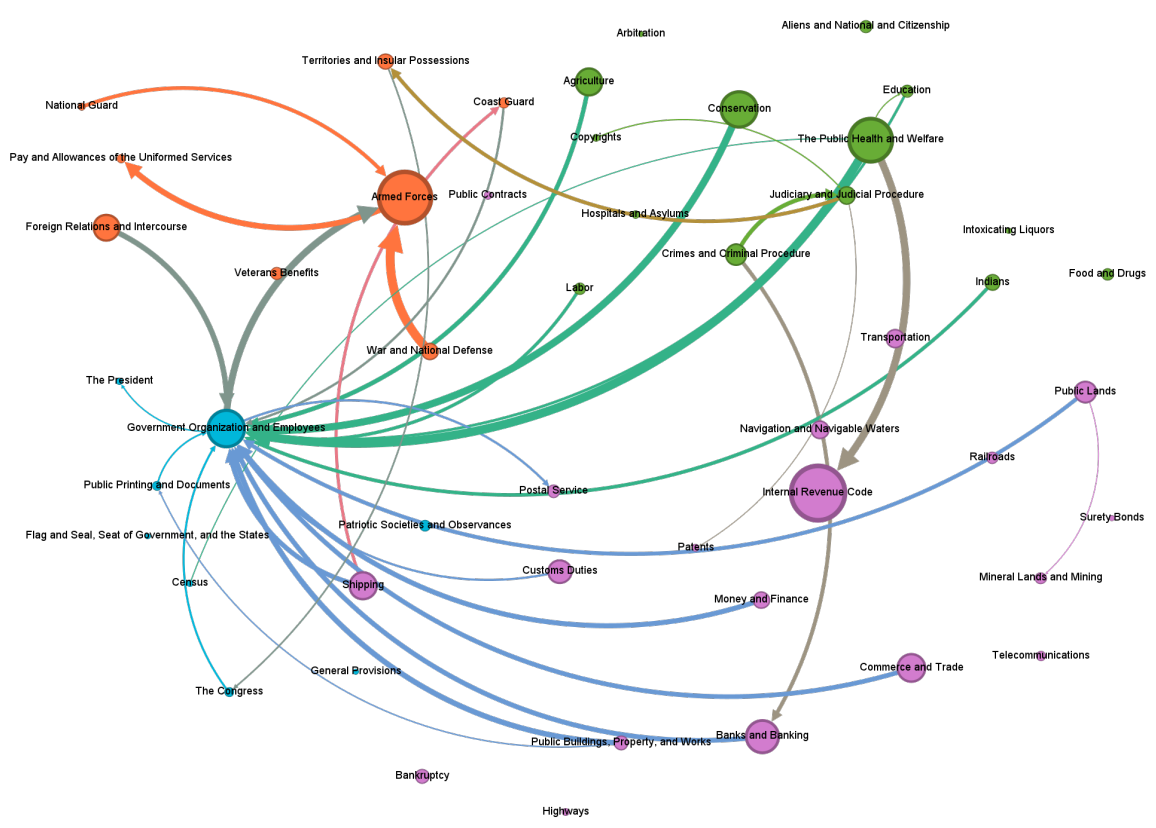

**(b)** year 1964

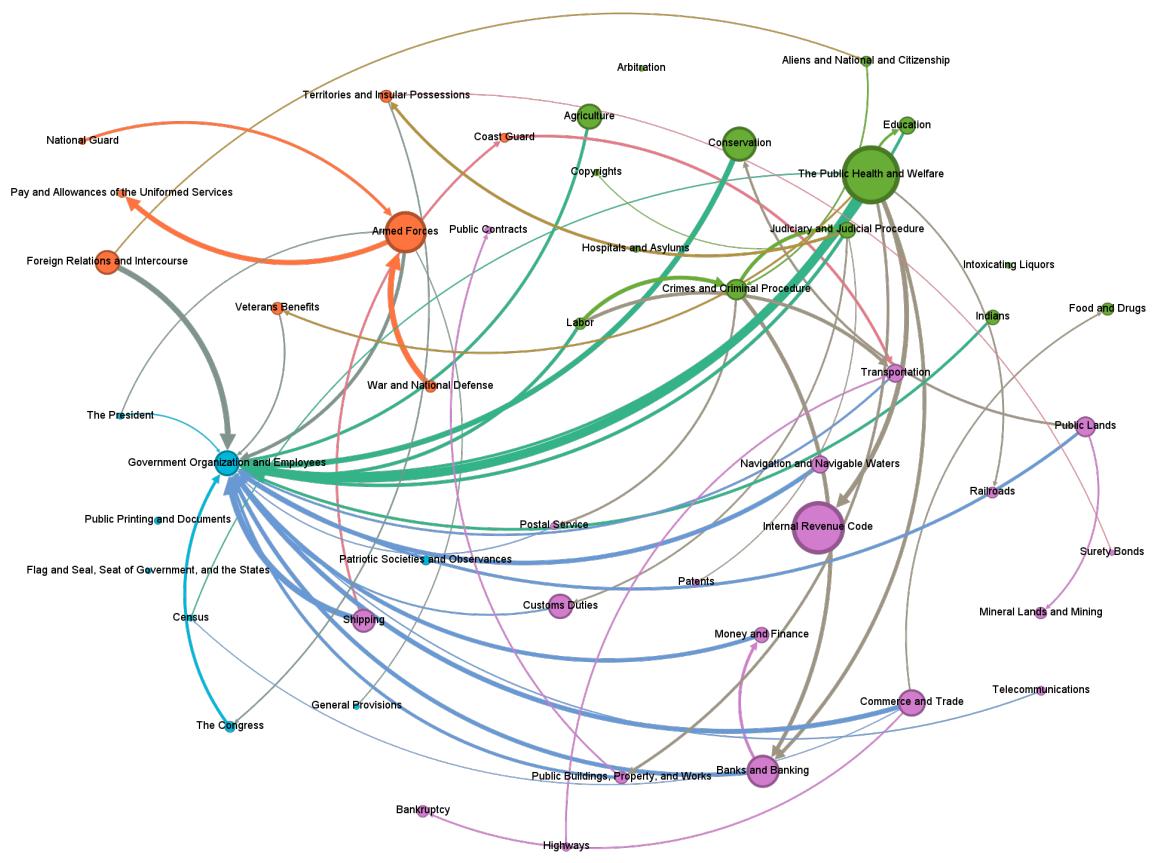

**(c)** year 1970

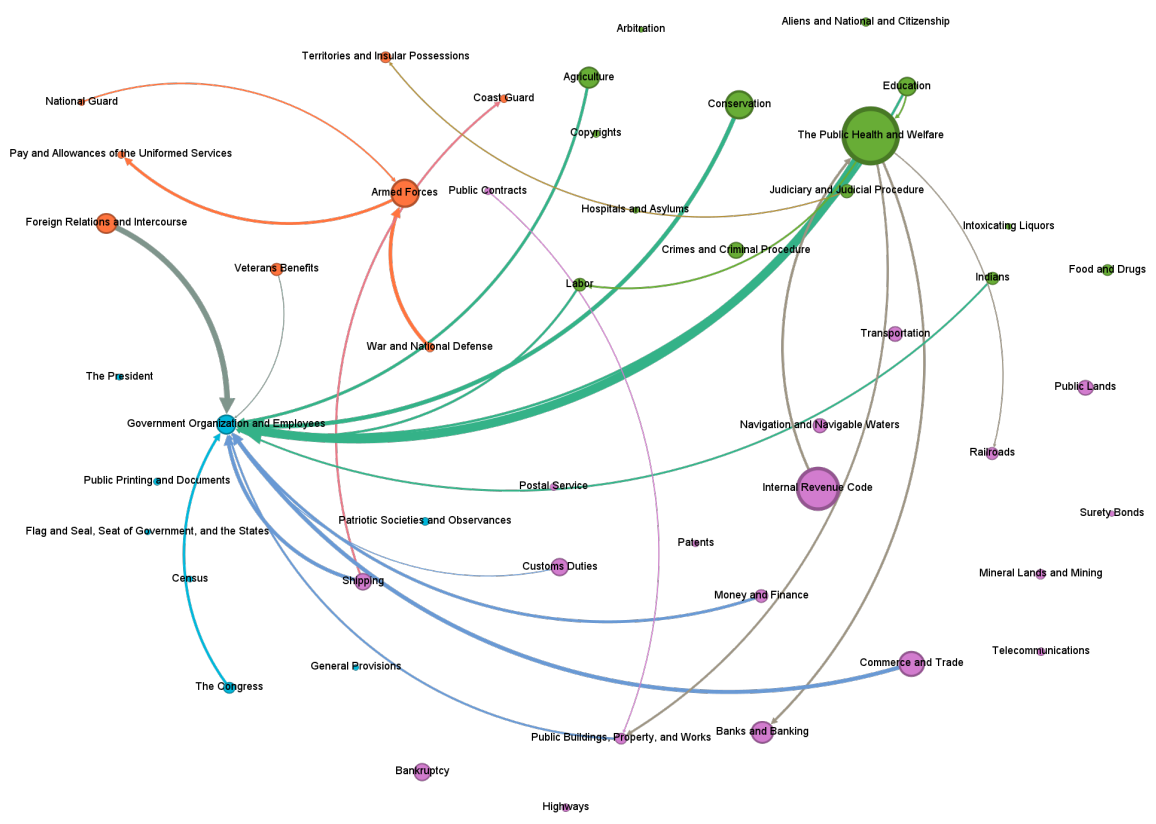

**(d)** year 1976

**Figure S1.12: Backbone structure of the U.S. Code's title cross-reference network in 1958, 1964, 1970, 1976.**

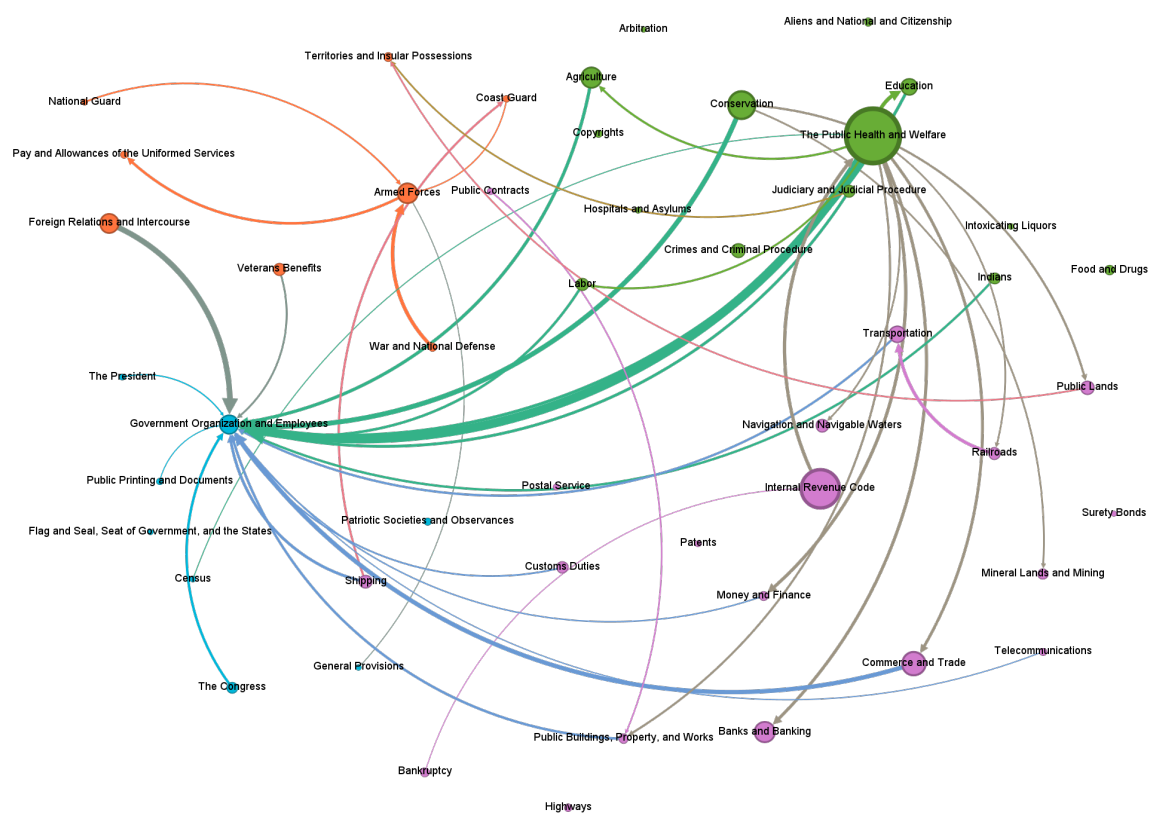

**(a)** year 1982

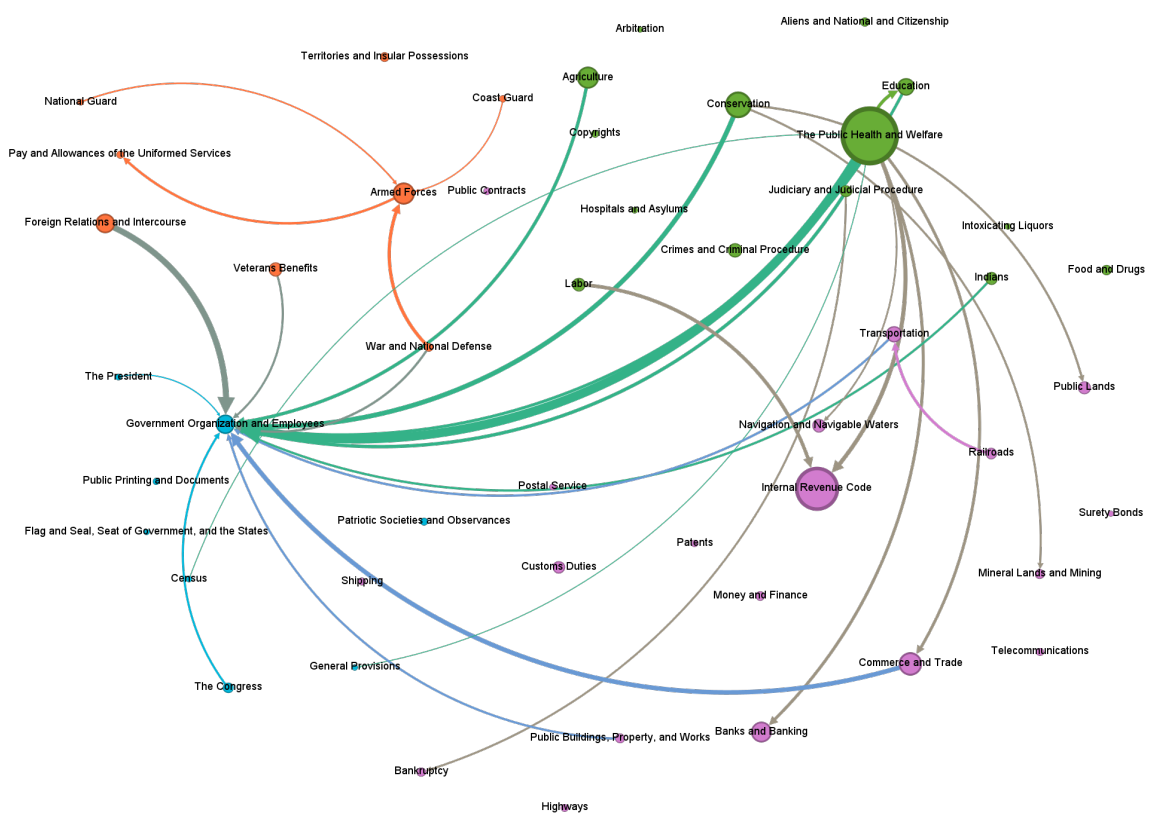

**(b)** year 1988

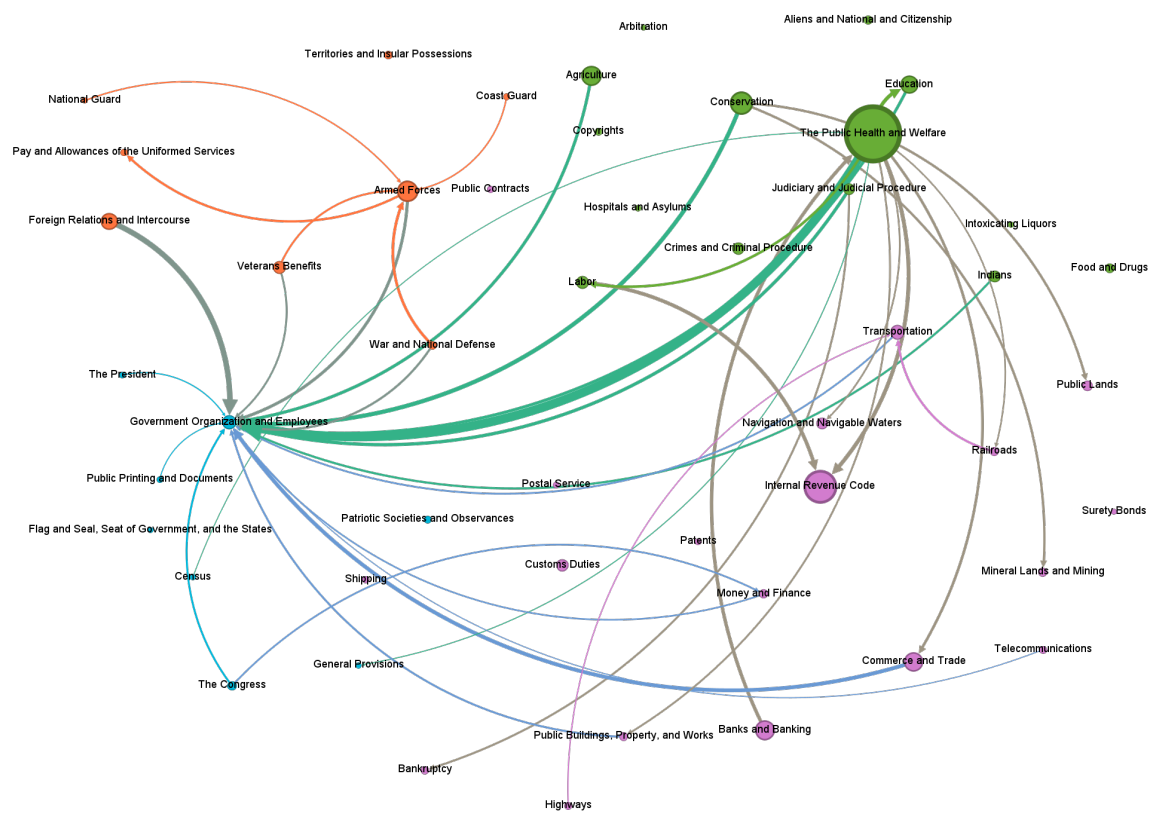

**(c)** year 1994

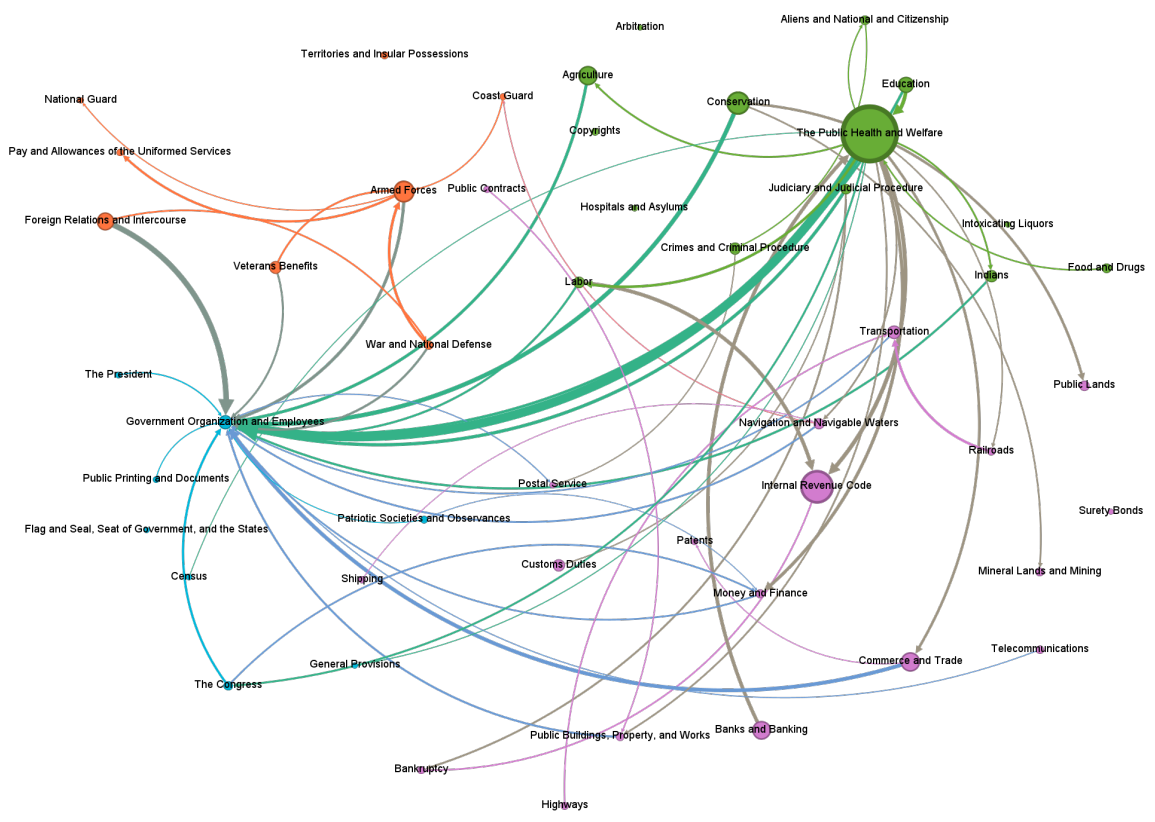

**(d)** year 2000

**Figure S1.13: Backbone structure of the U.S. Code's title cross-reference network in 1982, 1988, 1994, 2000.**

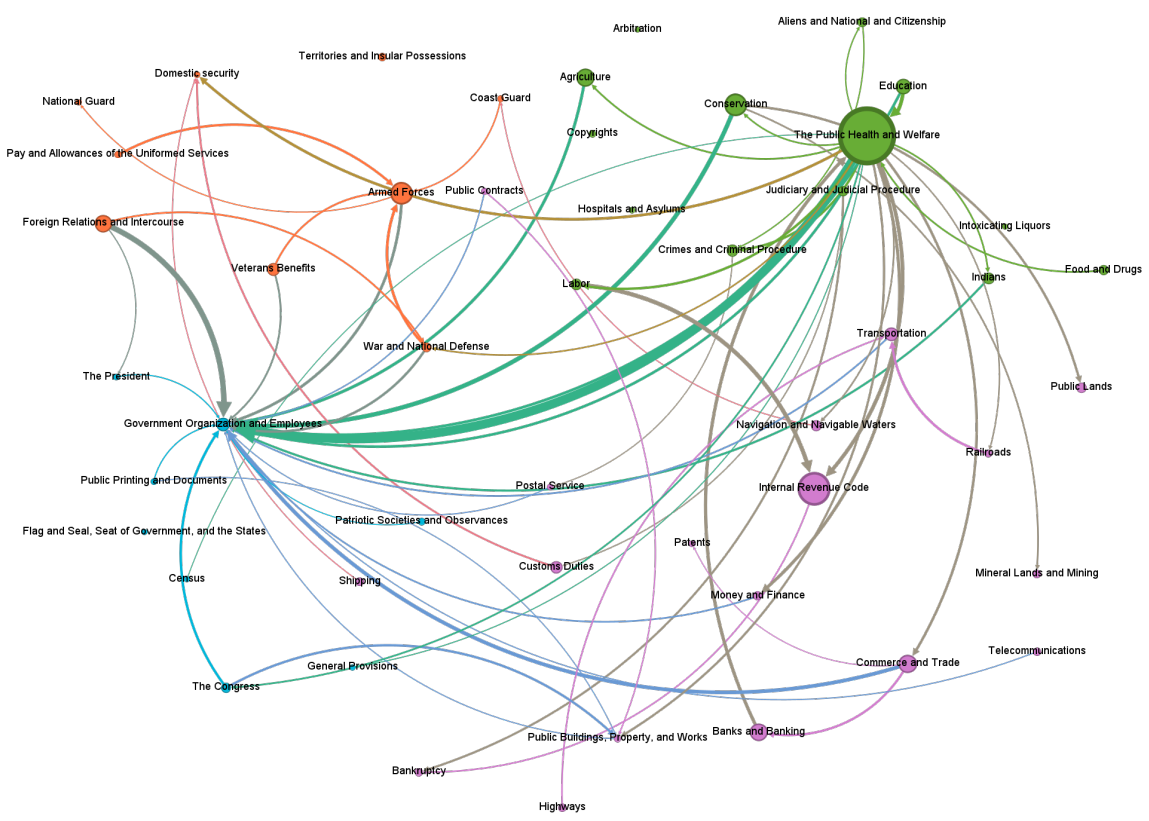

**(a)** year 2006

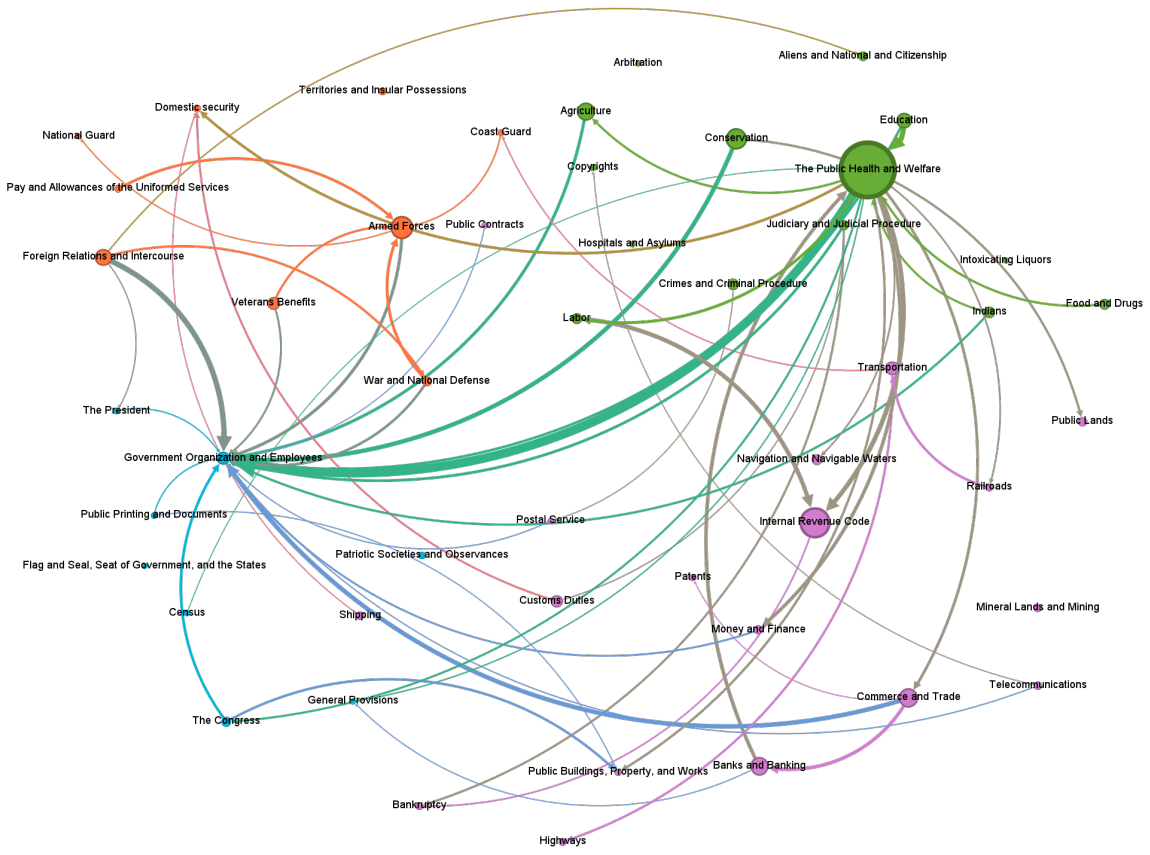

**(b)** year 2012

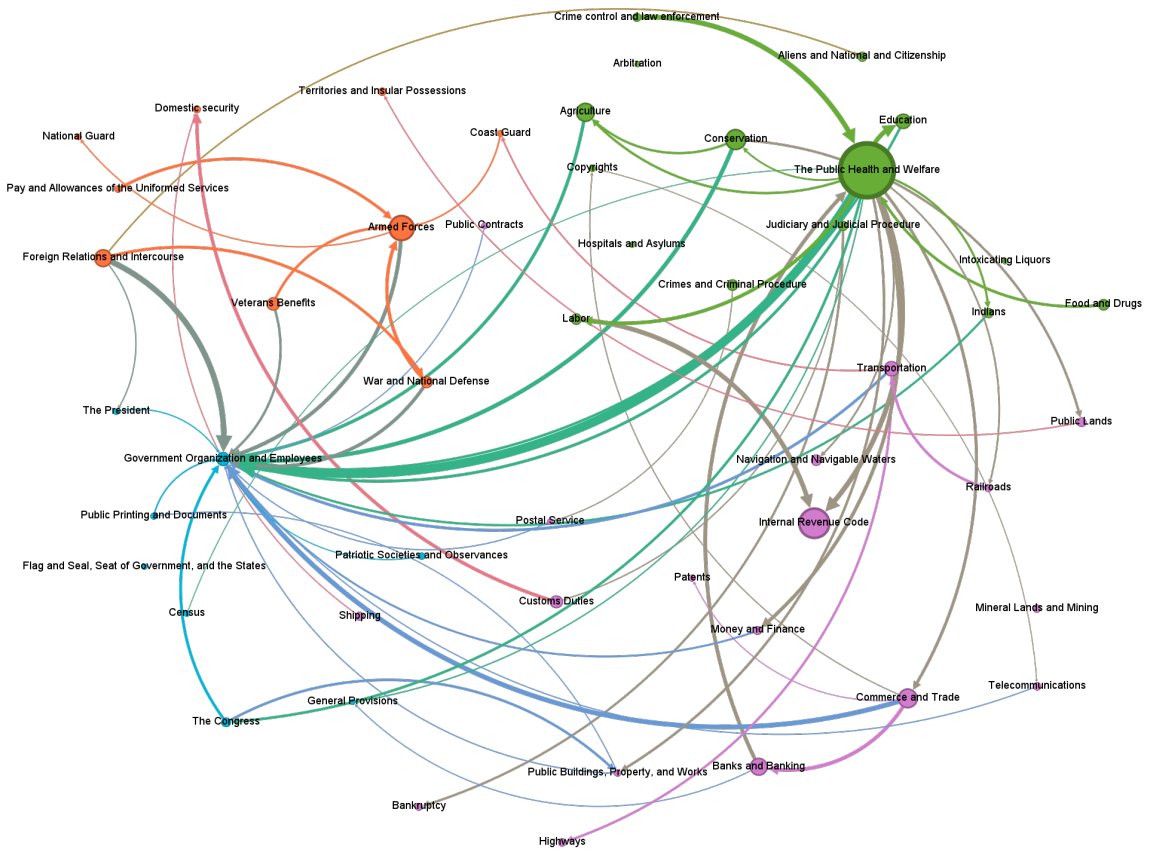

**(c)** year 2018

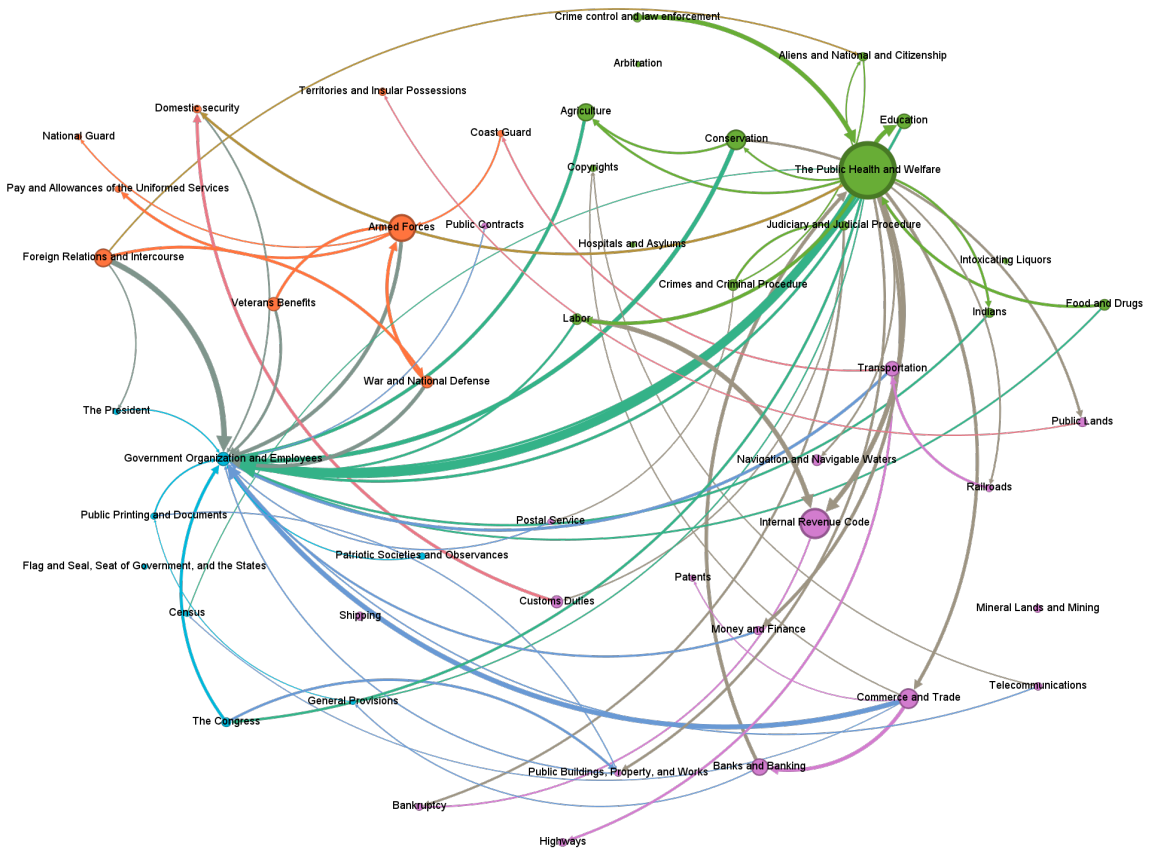

**(d)** year 2023

**Figure S1.14: Backbone structure of the U.S. Code's title cross-reference network in 2006, 2012, 2018, 2023.**